# RG-Based Local Hopf Reduction and Slow-Manifold Reconstruction for Nonlinear Aeroelastic Systems

Gelin Chen [1], Chen Song[2] and Chao Yang[3]

School of Aeronautic Science and Engineering, Beihang University, Beijing, China

**Abstract:** Self-excited limit-cycle oscillations (LCOs) from Hopf bifurcations are a key feature of nonlinear aeroelasticity and depend sensitively on structural and aerodynamic parameters. Classical center-manifold and normal-form theory describe this local behavior, but can be cumbersome to apply in large discretized models and standard reduced-order modeling (ROM) workflows. A renormalization-group (RG)-based reduction is developed that directly yields a Hopf-type amplitude equation on a local invariant manifold, specialized for polynomial nonlinearities in tensor-based discretizations and compatible with finite-element-type settings. The method provides explicit coefficients governing the Hopf threshold, criticality, and leading LCO amplitude/frequency trends, and admits a companion slow-manifold approximation with selected stable modes retained as static coordinates. Representative nonlinear-aeroelastic examples illustrate how the proposed framework supplies compact, parameter-aware Hopf/LCO descriptors suitable for local ROM construction near flutter.

---

[1]Gelin Chen, Doctoral Candidate, School of Aeronautics, Beihang University, E-mail:sy2405230@buaa.edu.cn.

[2]Chen Song, Associate Professor, School of Aeronautics, Beihang University, E-mail: songchen@buaa.edu.cn.

[3]Chao Yang, Associate Professor, School of Aeronautics, Beihang University, E-mail: yangchao@buaa.edu.cn.

## 1. Introduction

Aeroelasticity is a canonical fluid–structure interaction (FSI) problem in which the coupling between a flexible structure and a flowing fluid can produce a wide range of dynamic responses relevant to the stability and response of aircraft wings[1]. Classical aeroelastic analysis typically assumes linear structures and aerodynamics and thus emphasizes linear flutter prediction; yet extensive wind-tunnel and flight-test evidence shows phenomena not predictable from linear theory, notably self-excited limit-cycle oscillations (LCOs) and random-like responses[2]. LCOs are therefore an inherently nonlinear and critical aspect of aeroelastic stability assessment [3], and they may arise from diverse mechanisms. On the structural side, concentrated nonlinearities such as freeplay and cubic stiffness, as well as hysteresis/bilinear restoring forces, have long been recognized, while distributed geometric nonlinearities of plate-like structures can also contribute to LCO onset and saturation. On the aerodynamic side, nonlinearities associated with flow separation at high incidence and unsteady shock oscillations in transonic flow, as well as dynamic stall, are known LCO sources. Even in a simplified two-degree-of-freedom setting, nonlinear effects can change flutter characteristics in a disturbance-dependent manner: the linear flutter speed may remain unchanged for small disturbances but typically decreases for larger disturbances, whereas a cubic hardening spring can yield a limited-amplitude flutter persisting well above the linear flutter speed[4]. These observations shift the engineering focus from a binary "stable/unstable" boundary to post-critical dynamics, specifically the onset of a Hopf bifurcation and the resulting LCO amplitudes and frequencies[2][5]. In the present aeroelastic context, a Hopf bifurcation refers to the loss of stability of an equilibrium through a complex-conjugate eigenvalue pair crossing the imaginary axis, leading locally to the birth of small-amplitude self-excited oscillations and, under suitable nonlinear saturation, an LCO [5]. This is not merely a modeling detail: subcritical LCO can be “highly undesirable” because it may appear suddenly at large amplitudes and even below the flutter point, with responses that depend strongly on initial conditions, while nonlinearities can also induce aeroelastic “pathologies” for certain aircraft-store configurations and render linear control strategies unreliable in strongly nonlinear regimes [6]. Complementarily, the same self-sustained oscillations can be intentionally exploited—for example, wing LCO has the potential to supply continuous electrical power via energy harvesting, and changing structural/harvester parameters can alter the LCO intensity and the wind-speed range over which LCO occurs[7]. Consequently, engineering analysis tools are needed that can (i) determine whether a Hopf bifurcation occurs and classify the ensuing LCO (e.g., supercritical versus subcritical), and (ii) quantify how key structural and aerodynamic parameters affect the bifurcation threshold and the steady-state LCO amplitude and frequency.

Near the onset of an aeroelastic LCO, the governing equations can be viewed as a small perturbation of a linearly stable equilibrium in which one oscillatory mode becomes critical. Linearizing about the steady state yields a Jacobian whose spectrum separates into strongly stable directions and a weak, nearly neutral oscillatory direction associated with a complex-conjugate pair. In this regime, the engineering question is intrinsically local: one seeks a compact description of how the critical oscillation is born and saturates, and how its amplitude and frequency vary with parameters. Accordingly, local Hopf/LCO prediction reduces to obtaining a low-dimensional amplitude equation (for the slow evolution of the oscillation envelope) together with a small set of coefficients—often expressed through Lyapunov-type quantities—that determine the Hopf criticality (supercritical versus subcritical) and the leading-order trends in post-critical LCO amplitude and frequency.

In engineering workflows, reduced-order models are most often constructed by projecting the dynamics onto a fixed linear subspace because this is straightforward and computationally efficient for large discretized aeroelastic models, including finite-element-based ones. For linear aeroelasticity, the standard reduction is modal truncation: the motion is approximated on a low-dimensional modal subspace spanned by a few eigenmodes, and trajectories initialized in that subspace remain there because linear dynamics does not generate components outside it. For nonlinear problems, the same philosophy underlies widely used ROMs based on POD/Karhunen–Loève-type bases[8][9] and related SVD constructions[10], as well as linear-basis enrichments such as modal derivatives and dual modes designed to better capture nonlinear couplings. In aeroelasticity, this ROM-centric view is also common in nonlinear simulation frameworks that combine FE structural models with reduced aerodynamic descriptions (e.g., panel-type models), and then introduce structural ROMs to enable repeated analyses (e.g., the survey and representative implementations [11][12][13]. ). However, many practical nonlinear-aeroelastic ROM pipelines remain partly a posteriori[14] in the sense that nonlinear terms are inferred or tuned from response data (e.g., via harmonic-balance/FFT-type procedures) rather than derived in a fully parametric, first-principles manner. This can limit "simulation-light" bifurcation prediction and rapid parametric studies when one ultimately wants robust, local Hopf/LCO characterization—especially Hopf criticality and the sensitivity of the LCO amplitude and frequency to structural and aerodynamic parameters—in large-scale models.

Classical bifurcation theory provides a systematic route to exactly this local information via center-manifold reduction and normal forms. Geometrically, the linear modal subspace that is invariant in the linear model is replaced (locally) by a curved invariant manifold: nonlinear couplings generate out-of-subspace components, yet there exists an invariant surface in phase space such that once a trajectory lies on it, it stays on it for all time. For a Hopf bifurcation, the relevant invariant object is the center manifold, which is tangent to the critical eigenspace associated with the complex pair; restricting the dynamics to this manifold yields a low-dimensional reduced system that captures the local birth of LCO. Normal-form computations then simplify the reduced system to a canonical Hopf form whose coefficients classify the bifurcation and predict leading-order trends in the LCO amplitude and frequency. In principle, this is a complete mathematical recipe.

In practice for large discretized aeroelastic models, however, computing these Hopf coefficients “from first principles” typically requires explicit high-order expansions of the full-order vector field (or equivalent multilinear forms) and repeated solutions of associated invariance/solvability conditions. When the state dimension is large and when multiple structural/aerodynamic parameters are of interest, this algebraic overhead and implementation burden can become the dominant barrier, so practitioners often revert to time-marching/continuation or data-assisted identification to infer LCO trends.

This gap motivates a reduction route that targets the same local Hopf information — criticality, postcritical amplitude and frequency trends, and parameter dependence — while remaining algorithmic within tensor-based discretization workflows and amenable to finite-element implementation. Classical center-manifold and normal-form theory therefore provide, at least in principle, a complete route to the desired Hopf descriptors. The renormalization-group (RG) approach adopted here can be viewed as another realization of this route. In the formulation of Chen, Goldenfeld and Oono, the RG equation is obtained by absorbing secular terms of a naive perturbation expansion into slowly varying "renormalized" amplitudes, and the resulting reduced dynamics recovers within a single framework the familiar outputs of multiple scales, averaging, and related singular-perturbation techniques[15]. Subsequent work has clarified that, for weakly nonlinear oscillatory systems, the RG amplitude equations coincide to leading order with classical Poincaré-Birkhoff normal forms and provide long-time approximations on the same $O\left(\varepsilon^{-1}\right)$ time windows [16].

A complementary geometric picture interprets the RG construction as building an invariant manifold that organizes the slow motion of the critical variables. Ei, Fujii and Kunihiro showed that the perturbative RG procedure can be recast as an invariant-manifold/equation-of-envelope problem, with the RG variables serving as intrinsic coordinates on this manifold[17]. Chiba developed this viewpoint into a $C^1$ approximation theory: under mild assumptions, the family of RG approximate solutions defines an "RG vector field" that is close to the original vector field and inherits normally hyperbolic invariant manifolds and their stability properties[18][19]. He further demonstrated that the RG method unifies averaging, multiple-time-scale constructions, (hyper)normal forms and center-manifold reduction within a single perturbative framework [18].

Within this theoretical backdrop, RG has also been applied directly to Hopf bifurcations. Das, Banerjee and Bhattacharjee used the Chen-Goldenfeld-Oono algorithm to derive amplitude equations in two- and three-dimensional systems and to distinguish super- and sub-critical Hopf cases from the sign of the leading nonlinear coefficient, after removing the linear term, in benchmark models such as the Lorenz and Rössler systems[20]. Their results illustrate that RG-based amplitude equations can encode both the onset of Hopf and the leading post-critical amplitude trends in a form that is tightly tied to the underlying parameters and scaling choices-features we will exploit in the aeroelastic setting below.

Accordingly, a renormalization-group (RG) viewpoint is adopted for weakly nonlinear oscillations: secular terms generated by a naive perturbation expansion are absorbed into a slow evolution of the complex critical amplitude, yielding an amplitude equation that is, to the computed order, equivalent to the Hopf normal form on a local invariant manifold. Building on the RG--center-manifold perspective developed by Chiba[21], the present paper specializes that viewpoint to local Hopf/LCO analysis in nonlinear aeroelastic systems and formulates it in a way that is directly usable with polynomial/tensor representations. In particular, Appendix A organizes the reduction as an explicit computational procedure, while Appendix B gives a Hopf-oriented justification of the truncated RG manifold and reduced flow used here. The aim is not to claim a general replacement of classical center-manifold or normal-form theory, but rather to make precise, for the present setting, why the RG reduction provides a controlled local Hopf descriptor and how its predictions persist in the full system under standard nondegeneracy and normal-hyperbolicity assumptions.

The main contributions of this work are threefold.

First, an RG-based local Hopf reduction procedure is presented for polynomial nonlinear systems, yielding in direct algorithmic form the reduced complex amplitude equation together with the coefficients governing the Hopf threshold, criticality, and leading-order LCO amplitude/frequency trends. A companion slow-manifold approximation is also introduced, with the option of retaining selected stable coordinates as static modes for reconstruction.

Second, at the level needed for the present Hopf reduction, a theoretical justification is provided for the RG construction used in the paper. Building on earlier RG manifold results, we show that the truncated RG map defines an approximate local invariant manifold and reduced flow, and that, under suitable spectral-gap, normal-hyperbolicity, and nondegeneracy conditions, this approximate manifold is $C^r$ -close to a true invariant manifold. As a consequence, the Hopf point and its local criticality predicted by the RG-reduced system persist in the full system with errors controlled by the truncation order.

Third, the numerical study shows how these reduced coefficients can be used as informative local descriptors in nonlinear aeroelasticity. In particular, the results clarify that replacing the true coupled center eigenspace by an undamped structural modal surrogate can give qualitatively incorrect Hopf-sensitivity predictions, even when modal-shape similarity is high. They also provide a branchwise decomposition of the Hopf cubic coefficient, showing how different cubic stiffness mechanisms contribute differently along distinct flutter branches, and how some induced cubic effects depend on mediation by noncritical modes.

## 2. RG viewpoint for local Hopf reduction

To extract local Hopf/LCO descriptors in a form compatible with tensor-based workflows, including finite-element-type implementations, a renormalization-group (RG) viewpoint is adopted for weakly nonlinear oscillations. The central idea is to construct, from a naive perturbation expansion, a closed slow evolution law for the complex amplitudes of the critical (center) modes. Concretely, solutions are expanded about an arbitrary reference time $t_0$ in a small bookkeeping parameter $\varepsilon$, which produces at each order a linear inhomogeneous problem driven by lower-order terms. The forcing at a given order can be decomposed into sums of elementary contributions-exponentials associated with combinations of eigenvalues, multiplied by polynomials in the center amplitudes and parameters-so that resonant terms (those with the same exponential factor as a center eigenvalue) can be identified explicitly. These resonant coefficients are precisely the terms that would generate secular growth in a naive expansion, and in RG they are instead used to define amplitude equations for the center variables.

The amplitude equations are obtained by imposing a time-shift (envelope) condition expressing that the physical approximation should not depend on the arbitrary choice of reference time $t_0$ ,

$$\left.\frac{\partial x^{(N)}\left(t;t_0,a\left(t_0\right)\right)}{\partial t_0}\right|_{t_0=t}=0,$$

which removes the explicit $\left(t-t_0\right)$ -growth that limits naive expansions to $O(1/\varepsilon)$ time windows and yields a reduced vector field for the center amplitudes.

In the Hopf case, the RG procedure yields, after the chosen near-critical scaling and, if desired, a mild time reparametrization/normalization, a reduced complex-amplitude equation of the form:

$$\dot{C}=\sigma(\mu;\varepsilon)C+\gamma(\mu;\varepsilon)C\,|\,C\,|^2+O(\varepsilon^{r+1},|\,C\,|^5),$$

where C is the renormalized center amplitude and the coefficients are those produced by the RG reduction at the selected truncation order. Here the real part of $\sigma$ controls the local instability threshold, $Re(\gamma)$ determines the local saturation tendency and hence the supercritical/subcritical character, while $\Im(\sigma)$ and $Im(\gamma)$ encode the leading oscillation-frequency correction. To avoid notational ambiguity, we reserve the classical Hopf normal-form coefficients (e.g., the first Lyapunov coefficient and its standard normal-form parametrization) for the explicitly transformed canonical form; the relation between the RG coefficients used here and the classical Hopf normal-form coefficients is summarized in Appendix C; see also [5].

The remaining non-resonant contributions provide bounded corrections and define a reconstruction (embedding) map from center amplitudes (and parameters) back to the full state; evaluating the bounded stable-direction response at the reference time leads to an approximate slow-manifold relation, and selected stable coordinates may be retained explicitly as "static modes" when desired.

## 3. RG-based reduction algorithm for polynomial nonlinearities

### 3.1 Problem setting, parameter suspension, and scaling choices

An autonomous weakly nonlinear system is considered

$$\dot{x} = Ax + \varepsilon g(x,\mu), \quad g(x,\mu) = \sum_{k=1}^{p} \varepsilon^{k-1} g_k(x,\mu),$$

where $x \in \mathbb{R}^n, 0 < \varepsilon \ll 1$, and $\mu \in \mathbb{R}^q$ denotes a vector of design/operating parameters. Parameter dependence is handled by suspension, i.e., augmenting the state with $\dot{\mu} = 0$, which allows the reduced coefficients to be obtained as explicit functions of $\mu$ .Importantly, we do not assume that each $g_k$ corresponds to a single fixed polynomial degree. Instead, each $g_k$ may collect multiple multilinear tensor contributions, $g_k(x,\mu) = \sum_{\ell \geq 2} L_k^{(\ell)}(x,\ldots,x;\mu)$ where $L_k^{(\ell)}$ is a symmetric $\ell$ -linear mapping. This flexibility is convenient in engineering practice because the bookkeeping in $\varepsilon$ is often driven by a chosen scaling (state and/or parameter scalings) rather than by polynomial degree alone.

For Hopf bifurcation analysis, a common near-critical scaling is

$$x = \varepsilon \hat{x}, \quad \mu = \mu_0 + \varepsilon^2 \hat{\mu},$$

so that the post-critical amplitude and parameter detuning enter at comparable orders. In implementation, it is often advantageous to proceed in two stages: first identify/assemble a polynomial representation for the unscaled nonlinear term $f(x,\mu)$ in

$$\dot{x} = Ax + \varepsilon f(x,\mu),$$

(e.g., via analytic model evaluation or static/quasi-static tests), and then impose the chosen scalings and redistribute contributions into the $\{g_k\}$ hierarchy. This "identify first, scale second" workflow makes it straightforward to change the $\varepsilon$ ordering without recomputing the tensor ingredients.

The matrix $A$ is assumed to be diagonalizable over $\mathbb{C}$ and admits a center-stable spectral split; for a simple Hopf point the center spectrum consists of one complex-conjugate pair $\lambda_c = \mathrm{i}\omega, \bar{\lambda}_c = -\mathrm{i}\omega$, and all other eigenvalues are stable. Let $U = [U_c \quad U_s]$ and $V = [V_c \quad V_s]$ collect right/left eigenvectors scaled such that

$$V_c^* U_c = I, \quad V_c^* U_s = 0, \quad V_s^* U_c = 0, \quad V_s^* U_s = I.$$

Modal coordinates $y = V^* x$ (so $x = Uy$) are used, but in practice only ($U_c$, $V_c$) and an optional user-selected subset of stable modes are required explicitly. Inputs. (i) $A$ and the Hopf pair $(\lambda_c, U_c, V_c)$ at the reference equilibrium; (ii) multilinear tensors needed by the chosen truncation order (typically up to cubic); (iii) an optional index set $\mathcal{I}_{\mathrm{sm}}$ of stable coordinates retained explicitly as static modes for reconstruction. Outputs. (i) an RG reduced equation for the center complex amplitudes $C(t)$ with coefficients depending on $\mu$; (ii) an embedding/reconstruction map $x \approx \Phi(C, \mu)$ (optionally enriched by static modes).Cost. Up to cubic order, the method requires only a small number of tensor contractions (e.g., quadratic/cubic forms tested on $U_c, \bar{U}_c$ and selected stable modes) and a small number of linear solves with shifted operators; no long-time time-domain integration is needed.

### 3.2 Perturbation hierarchy and the forcing structure $G_i$

A perturbation expansion is introduced

$$x = x_0 + \varepsilon x_1 + \varepsilon^2 x_2 + \cdots,$$

substitute it into $\dot{x} = Ax + \varepsilon g(x, \mu)$, and collect like powers of $\varepsilon$. This yields the perturbation

hierarchy

$$\dot{x}_0 = Ax_0, \quad \dot{x}_i = Ax_i + G_i\left(x_0,\dots,x_{i-1};\mu\right), \quad i \ge 1,$$

where the forcing terms $G_i$ are built algebraically from $\{g_k\}$ and derivatives evaluated along lower-order solutions. For the first few orders (shown here to make the structure explicit), one obtains

$$G_1 = g_1\left(x_0\right), \quad G_2 = Dg_1\left(x_0\right)x_1 + g_2\left(x_0\right),$$
$$G_3 = Dg_1\left(x_0\right)x_2 + \frac{1}{2}D^2 g_1\left(x_0\right)\left(x_1,x_1\right) + Dg_2\left(x_0\right)x_1 + g_3\left(x_0\right),$$

and higher-order $G_i$ follow the same pattern (see Appendix A for the general expression). The key point for implementation is that, under the polynomial-tensor assumption, the class of expressions appearing in $G_i$ is closed under differentiation, so the assembly of each $G_i$ reduces to finite tensor contractions and additions. Center-only choice at leading order. At $\varepsilon^0$ the solution $x_0\left(t\right)$ is chosen to lie purely in the center subspace,

$$x_0(t) = X(t)U_c C, \quad X(t) = e^{At},$$

with $C$ collecting the complex amplitudes (with the usual conjugacy constraint for real $x$ ). This choice eliminates decaying stable transients from the outset and ensures that resonance bookkeeping at higher orders is governed solely by center frequencies (as in Appendix A).

### 3.3 Elementary-term representation and resonance (Hopf setting)

For the Hopf case with $x_0$ center-only, every forcing contribution in $G_i$ can be decomposed into a finite sum of elementary terms

$$G(t;C,\overline{C},\mu) = p(C,\overline{C},\mu)e^{\lambda t}U_j,$$

where $U_j$ is a (center or stable) eigenvector of $A, \lambda$ belongs to the integer module generated by the center eigenvalues (for a simple Hopf point: integer multiples of $\pm i\omega$ ), and $p(\cdot)$ is a polynomial in $(C,\bar{C},\mu)$ . These coefficient polynomials are written generically as $p$ to emphasize that, once tensors are available, the construction is purely algebraic (Appendix A. 5 gives the closure argument). For an elementary term $pe^{\lambda t}U_j$ , the Duhamel integral that produces a particular solution has two qualitatively distinct outcomes:

- Resonant case ( $\lambda=\lambda_j$ ): the integral generates a secular factor $tpe^{\lambda_j t}U_j$ .

- Nonresonant case ( $\lambda\neq\lambda_j$ ): the integral yields a bounded oscillatory correction proportional to $\left(\lambda-\lambda_j\right)^{-1}pe^{\lambda t}U_j$ .

These two cases are the only mechanism needed for the RG bookkeeping (cf. Appendix A.5). In particular, because $x_0$ contains only purely imaginary center exponents, the exponents $\lambda$ appearing in elementary terms have no real part; therefore the stable eigendirections do not generate secular terms, and resonant contributions arise only in the center eigendirections. This is precisely why the RG secular terms are defined by projection onto $V_c$ in Appendix A.

### 3.4 RG definitions of secular terms $\mathbf{R}_\mathbf{i}$ and bounded corrections $h^{(i)}$

Following Chiba's formulation, At each order, an RG secular contribution is defined $R_i(C,\bar{C},\mu)$ (governing the slow evolution of the center amplitudes) and a bounded correction $h_t^{(i)}(C,\bar{C},\mu)$ (contributing to reconstruction). For the first order,

$$R_1(C)=\lim_{t\to\infty}\frac{1}{t}\int_0^t V_c^* X^{-1}(s)g_1\left(X(s)U_cC\right)ds$$

$$h_t^{(1)}(C)=\int^t X(t)X^{-1}(s)\left[g_1\left(X(s)U_cC\right)-X(s)U_cR_1(C)\right]ds$$

where the (unspecified) lower integration limit indicates that a particular solution is chosen to preserve the elementary-term structure (Appendix A.4-A.5).

For higher orders, the same separation is applied after assembling the forcing $G_i$ and subtracting the pieces induced by the slow evolution of $C(t)$. One convenient recursive definition is (Appendix A.4)

$$R_i(C)=\lim_{t\to\infty}\frac{1}{t}\int_0^t V_c^* X^{-1}(s)\left[G_i\left(X(s)U_c C, h_s^{(1)}(C),\ldots,h_s^{(i-1)}(C)\right)-\sum_{k=1}^{i-1} Dh_s^{(k)}(C)R_{i-k}(C)\right]ds$$

with the associated bounded correction

$$h_t^{(i)}(C)=\int^t X(t)X^{-1}(s)\left[G_i(\cdots)-\sum_{k=1}^{i-1} Dh_s^{(k)}(C)R_{i-k}(C)-X(s)U_c R_i(C)\right]ds$$

Where the "chain-rule" terms come from (and why they matter). The terms $Dh_s^{(k)}(C)R_{i-k}(C)$ are not an implementation nuisance but a structural necessity: the envelope (reference-time invariance) condition differentiates the entire perturbation solution with respect to the reference time, so not only secular terms $t\,p(C)$ but also bounded nonresonant terms contribute through their $C$ -dependence. Those contributions appear at higher orders as effective nonautonomous forcing unless they are systematically accounted for. The recursive subtraction above exactly captures this effect, while remaining purely algebraic once the elementary-term representation is available. After truncation at order $m$, the RG reduced equation and reconstruction map read

$$\dot{C}=\sum_{i=1}^{m}\varepsilon^i R_i(C,\bar{C},\mu),\quad \Phi_t(C)=X(t)U_c C+\sum_{i=1}^{m}\varepsilon^i h_t^{(i)}(C,\bar{C},\mu).$$

**3.5 Slow-manifold approximation and static-mode enrichment**

The bounded corrections $h_t^{(i)}$ contain both center and stable components. A practical slow-manifold approximation is obtained by evaluating the bounded part at a chosen reference time (or, equivalently, dropping the homogeneous stable transients by construction), giving an algebraic slaving relation for the stable coordinates,

$$y_s \approx h_s(C, \bar{C}, \mu).$$

This provides an explicit local embedding map $x \approx \Phi(C,\mu)$ that can be used to reconstruct physical states (e.g., structural displacements and aerodynamic states) from the center amplitudes.

When desired, a selected subset of stable coordinates can be retained explicitly as static modes $y_{\mathrm{sm}}$ (indexed by $\mathcal{I}_{\mathrm{sm}}$), while the remaining stable coordinates are still enslaved. This option is useful when certain stable directions have strong quadratic/cubic coupling into the slow manifold and one wishes to expose their contribution in the reconstruction without increasing the dimension of the dynamical core (the amplitude equation remains center-based).

**3.6 Hopf normal form and required tensor contractions (summary)**

For a simple Hopf pair, the RG reduced equation can be written in the familiar complex normal-form style

$$\dot{C} = \lambda(\mu)C + a(\mu)|C|^2 C + b(\mu)|C|^4 C + \mathcal{O}\left(|C|^6\right),$$

where $\lambda(\mu) = \sigma(\mu) + \mathrm{i}\omega(\mu)$ and $a(\mu)$ determines criticality and leading-order post-critical trends. In computation, these coefficients reduce to a finite set of multilinear tensor contractions (quadratic/cubic tensors evaluated on $U_c, \bar{U}_c$ and, when included, on static/enslaved stable responses) and a small number of shifted linear solves/resolvent actions. Exact coefficient formulas and the theoretical justification of the RG reduction for Hopf points are provided in Appendix B.

The relation between the RG coefficients and the classical center-manifold/normal-form coefficients is summarized in Appendix C.

## 4. Application Example

### 4.1 Problem formulation and RG-oriented local parameterization

As an application example, a three-degree-of-freedom airfoil section is considered with plunge $h$, pitch $\alpha$, and control-surface deflection $\beta$, together with an aerodynamic augmented variable $z$, as shown in Fig. 1. The variable $z$ is introduced to represent the memory effect of unsteady aerodynamics in the time domain. More specifically, the aerodynamic loads are described by Theodorsen-type unsteady aerodynamic theory, and the corresponding frequency-domain representation is converted into an equivalent finite-dimensional time-domain model through a Wagner-function-based approximation. This leads to an augmented aeroelastic system that retains the essential unsteady aerodynamic effect while remaining suitable for local bifurcation analysis. Structural damping is modeled by Rayleigh damping.

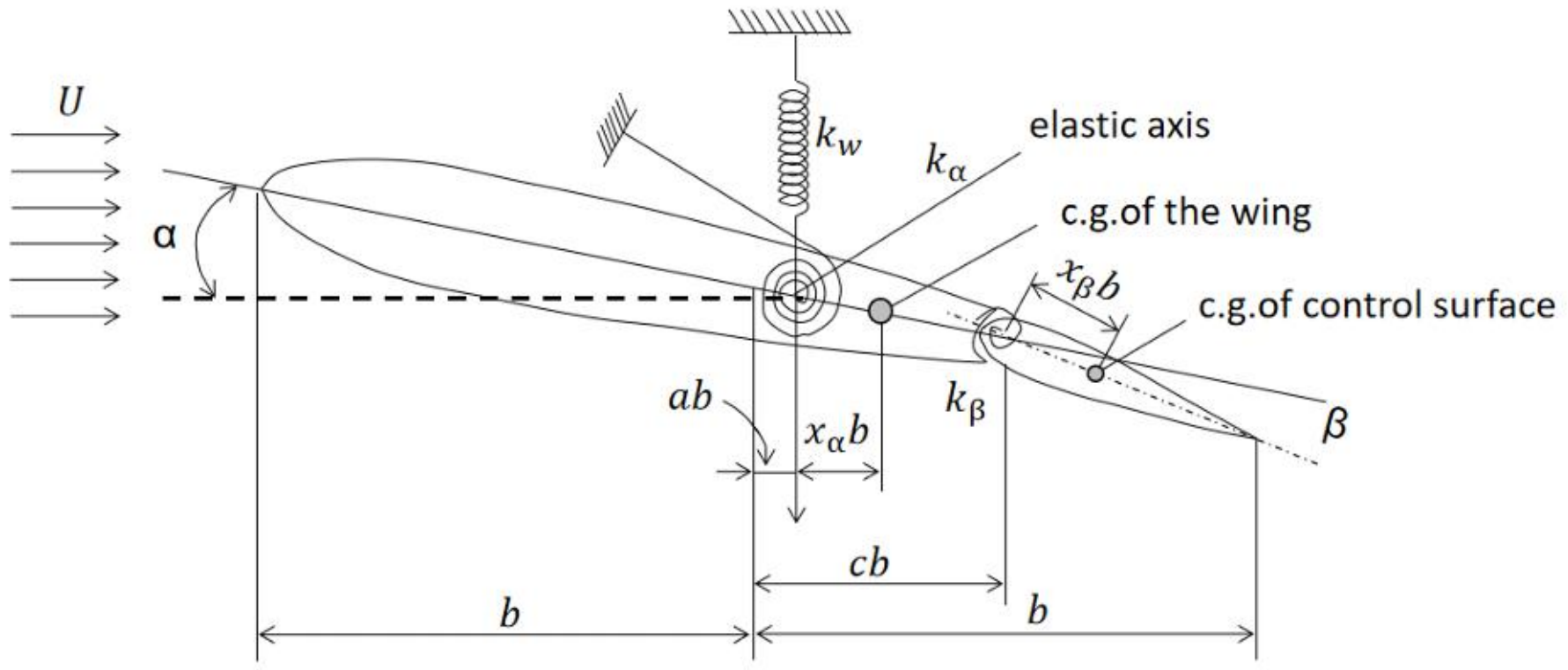

**Fig. 1** Schematic of the three-degree-of-freedom airfoil section with control surface.

Let the generalized coordinate vector be defined as $q = [h, \alpha, \beta, z]^T$. With this notation, the coupled system is written schematically in the second-order form

$$M\left(U,x_a\right)\ddot{q}+C\left(U,x_a\right)\dot{q}+K\left(U,x_a,k_{\beta 0}\right)q+F_{\mathrm{nl}}(q)=0$$

where $M$, $C$, and $K$ denote the effective mass, damping, and linear stiffness matrices of the coupled aeroelastic system, respectively. Their entries depend on the flow speed $U$ and the dimensionless distance between the elastic axis and the center of mass denoted by $x_\alpha$. Specific matrix constructions are provided in Appendix D, with parameter values given in Table 1. The quantity $k_{\beta 0}$ denotes the control-surface linear stiffness and is used below as the principal unfolding parameter in the local Hopf analysis. The nonlinear term $F_{\mathrm{nl}}(q)$ collects the structural restoring nonlinearities considered in this work.

**Table 1** Parameter values of the system

| *Parameter* | *Mean* | *Value* |
|---|---|---|
| $b$ | Half-chord length | 0.127m |
| $a$ | Dimensionless distance of elastic axis from mid-chord | -0.6 |
| $c$ | Dimensionless distance of hinge point from mid-chord | 0.6 |
| $\rho$ | Air density | $1.227kg/m^3$ |
| $m_T$ | Total mass per unit length of wing and added mass | $1.716+1.814kg/m$ |
| $\omega_w$ | Uncoupled plunge natural frequency | 41.5211rad/s |
| $\omega_\alpha$ | Uncoupled pitch natural frequency | 48.7963 rad/s |
| $\omega_\beta$ | Uncoupled control surface natural frequency | 116.5226 rad/s |
| $x_\beta$ | Dimensionless distance of control surface center of mass from hinge | 0.01795 |
| $r_\alpha$ | Dimensionless wing radius of gyration about the elastic axis | 0.6840 |
| $r_\beta$ | Dimensionless control surface radius of gyration about the elastic axis | 0.07336 |
| $\zeta_\beta$ | Plunge damping ratio | 0.015 |
| $\zeta_\alpha$ | Pitch damping ratio | 0.01 |
| $\zeta_w$ | Control surface damping ratio | 0.012 |
| $k_{h0}, k_{h3}$ | Plunge stiffness coefficients | 1 |
| $k_{\alpha 0}, k_{\alpha 3}$ | Pitch stiffness coefficients | 1 |
| $k_{\beta 0}, k_{\beta 3}$ | Control surface stiffness coefficients | 1 |

In the cubic-stiffness setting adopted in the numerical examples, the nonlinear restoring force is decomposed into three contributions associated with plunge, pitch, and control-surface nonlinear stiffness, namely

$$F_{\rm nl}(q) = k_{h3}F_h^{(3)}(q) + k_{\alpha 3}F_\alpha^{(3)}(q) + k_{\beta 3}F_\beta^{(3)}(q),$$

Where $k_{h3}, k_{\alpha 3}$ , and $k_{\beta 3}$ are the corresponding cubic stiffness coefficients, see D.2. Each contribution is homogeneous of degree three in the generalized coordinates. Equivalently, these terms may be represented by unsymmetrized cubic tensors in the full state space. This tensor representation is useful for the implementation of the RG coefficients, although the explicit tensor entries are not needed in the main text.

For the RG reduction and the subsequent modal interpretation, the second-order system is rewritten in first-order form. Defining the state vector as

$$X = [h, \alpha, \beta, z, \dot{h}, \dot{\alpha}, \dot{\beta}, \dot{z}]^T,$$

the governing equation becomes

$$\dot{X} = A\left(U, x_a, k_{\beta 0}\right)X + f_{\mathrm{nl}}\left(X; k_{h3}, k_{\alpha 3}, k_{\beta 3}\right),$$

where the linear operator has the standard block form

$$A\left(U, x_a, k_{\beta 0}\right) = \begin{bmatrix} 0 & I \\ -M^{-1}K & -M^{-1}C \end{bmatrix},$$

and the nonlinear vector field is

$$f_{\mathrm{nl}}(X) = \begin{bmatrix} 0 \\ -M^{-1}F_{\mathrm{nl}}(q) \end{bmatrix}.$$

Although the full state contains the aerodynamic augmented coordinate $z$ and its velocity $\dot{z}$, the nonlinearity in the present model is purely structural and enters only through the restoring-force term.

Accordingly, the first-order nonlinear vector field can be written as

$$f_{\mathrm{nl}}(X) = k_{h3} f_h^{(3)}(X) + k_{\alpha 3} f_\alpha^{(3)}(X) + k_{\beta 3} f_\beta^{(3)}(X),$$

where each $f_{\bullet}^{(3)}(X)$ is homogeneous of degree three. This decomposition is important because the reduced cubic coefficient near Hopf inherits the same additivity with respect to the individual cubic stiffness mechanisms.

To connect the full augmented system with the local RG reduction, we adopt the standard

near-critical scaling

$$k_{\beta 0} = k_{\beta 0}^{\star} + \varepsilon^2 \delta k_{\beta 0}, \quad X = \varepsilon Y,$$

where $Y$ is an $O(1)$ scaled state vector. Under this scaling, the linear detuning caused by the perturbation in $k_{\beta 0}$ and the leading cubic nonlinear effects appear at the same asymptotic order. Expanding the system about the reference Hopf point yields

$$\dot{Y} = A_0 Y + \varepsilon^2 \left( A_\beta \delta k_{\beta 0} Y + f_{\mathrm{nl}}^{(3)}(Y) \right) + O\left(\varepsilon^4\right),$$

where $A_0 = A\left(U_0, x_{a0}, k_{\beta 0}^{\star}\right), A_\beta = \partial A / \partial k_{\beta 0}\big|_{\left(U_0, x_{a0}, k_{\beta 0}^{\star}\right)}$, and

$$f_{\mathrm{nl}}^{(3)}(Y) = k_{h3} f_h^{(3)}(Y) + k_{\alpha 3} f_\alpha^{(3)}(Y) + k_{\beta 3} f_\beta^{(3)}(Y).$$

This is the form used for the RG reduction. In particular, the cubic stiffness coefficients are treated as $O(1)$ quantities under the present scaling, so their induced nonlinear effects enter the reduced equation at the same order as the parameter dependent linear detuning. As a consequence, the cubic coefficient in the reduced Hopf equation can be decomposed naturally into contributions associated with $k_{h3}$, $k_{\alpha 3}$, and $k_{\beta 3}$.

For the parameter range considered here, the linearized coupled system admits two Hopf bifurcation branches in the $(U, x_\alpha)$ plane, corresponding to two distinct flutter mechanisms; see Fig. 2. In what follows, these are referred to as the bending-dominated flutter branch and the control-surface-dominated flutter branch, according to the modal character of the associated critical eigenspace.

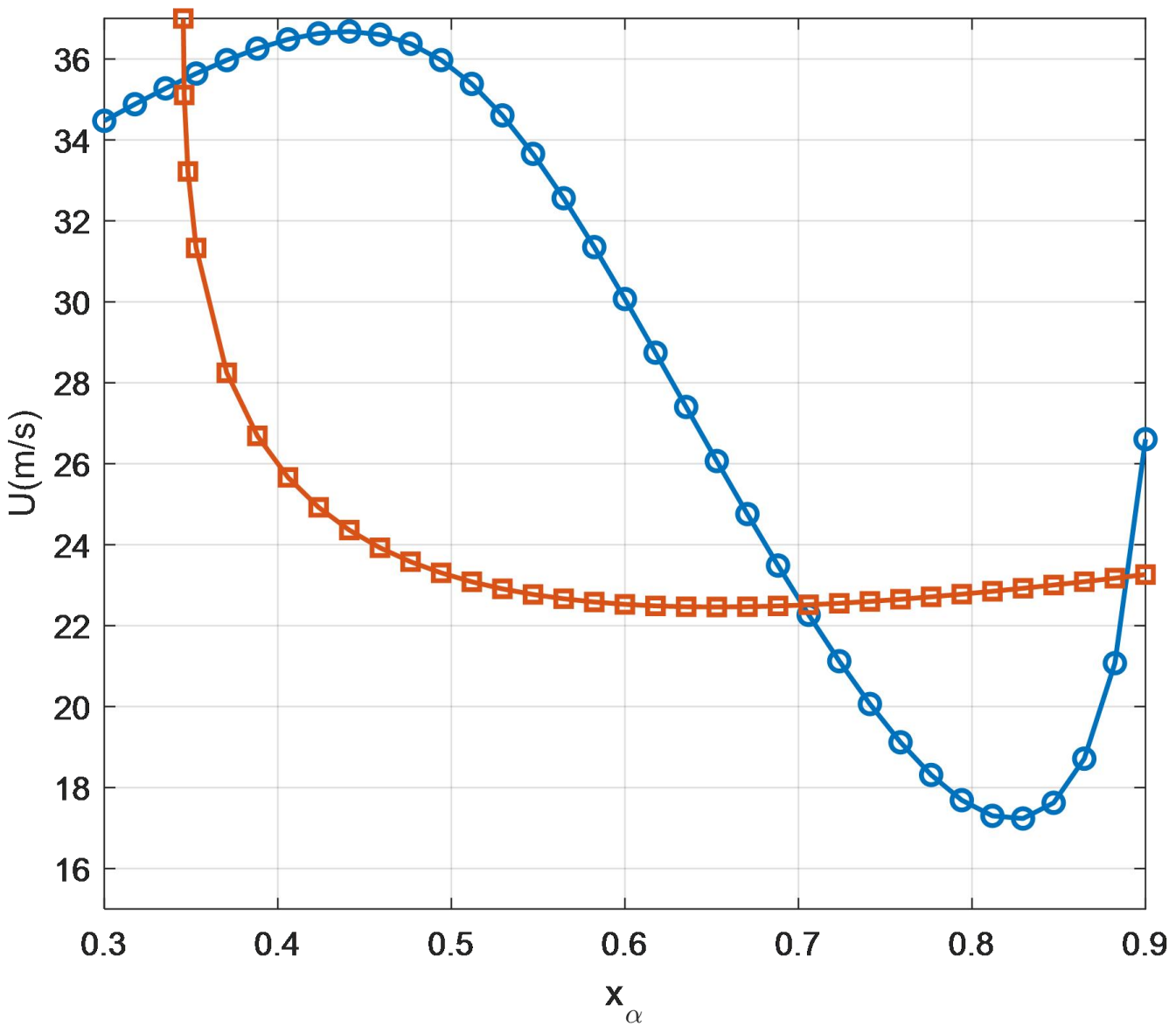

**Fig. 2** Hopf bifurcation curves in the $U - x_\alpha$ parameter plane. Branch 1(blue): control-surface flutter; Branch 2(red): bending-dominated flutter.

Near a selected point on either branch, the RG procedure yields a reduced amplitude equation for the critical complex modal coordinate $C$ of the form

$$\dot{C} = \varepsilon^2 \sigma C + \varepsilon^2 \gamma |C|^2 C + \text{ higher-order terms },$$

where $\sigma$ is the linear unfolding coefficient and $\gamma$ is the cubic coefficient governing the leading nonlinear correction to the Hopf bifurcation. Throughout this paper, $\gamma$ is referred to as the Hopf cubic coefficient. This terminology is used deliberately: it emphasizes the role of $\gamma$ in determining the local Hopf criticality while avoiding an unnecessary identification with a convention-dependent definition of the first Lyapunov coefficient.

Two features of this reduced equation are especially important for the numerical study. First, the coefficient $\sigma$ is not an empirical fitting parameter; it is obtained by projecting the linear perturbation of the augmented operator onto the critical eigenspace, and it may therefore be interpreted directly as the local sensitivity of the critical eigenvalue to the parameter $k_{\beta 0}$ . Second, because the cubic vector field enters linearly through the nonlinear tensor, the Hopf cubic coefficient $\gamma$ inherits an additive decomposition with respect to $k_{h3}$ , $k_{\alpha 3}$ , and $k_{\beta 3}$ . This makes it possible to compare the effects of different nonlinear stiffness components cleanly along the two Hopf branches.

These observations provide the basis for the numerical study below. Rather than treating the RG equation as a purely formal reduced model, Its coefficients are used here as local descriptors of two practically relevant features of the nonlinear aeroelastic system: the drift of the critical eigenvalue under parameter perturbation, and the branch-dependent tendency toward supercritical or subcritical behavior near the flutter boundary.

**4.2 Numerical study of the RG-based local Hopf/LCO descriptors**

**4.2.1 local prediction of subcritical behavior and unstable limit cycle**

First, it is examined whether the RG reduction can provide a quantitatively useful local description of a subcritical Hopf bifurcation in the nonlinear aeroelastic system. The parameter set considered in this discussion is

$$\varepsilon = 1, \quad \delta k_{\beta 0} = -0.05, \quad k_{\beta 3} = 45, \quad U_0 = 26.4322, \quad x_{a0} = 0 \qquad 6$$

for which the reduced equation predicts a subcritical-type local structure. At the selected Hopf point, the RG equation truncated to the order used in the computation yields an effective amplitude dynamics of the form

$$\dot{C} = \sigma_{\text{eff}} C + \gamma_{\text{eff}} |C|^2 C + \cdots,$$

with

$$\sigma_{\text{cff}} = -0.0913 - 0.0408i, \quad \gamma_{\text{eff}} = 0.0709 + 0.0326i.$$

The real parts satisfy

$$\Re(\sigma_{\text{eff}}) < 0, \quad \Re(\gamma_{\text{eff}}) > 0,$$

which implies a subcritical local Hopf tendency in the reduced dynamics: the equilibrium is still locally attracting on the center manifold, but an unstable finite-amplitude cycle exists at a nonzero radius. The radius predicted from the reduced equation is

$$r_{\text{RG}} = 1.13487.$$

This radius should be interpreted as a local center-manifold threshold estimate, rather than as a global basin boundary of the full system.

To test this local prediction, two trajectories were initialized along the critical modal direction, one slightly inside and the other slightly outside the predicted threshold, with initial radii 1.0781 and 1.3051 , respectively. The resulting time-domain responses are shown in Fig. 3. For the inside case, the full-system response remains attracted to the equilibrium, and the zoomed view shows that the reduced and full responses remain in close agreement over the highlighted local window. For the outside case, the trajectory leaves the local neighborhood and escapes after a finite time. The enlarged view indicates that visible disagreement begins at about $t \approx 8.2s$ , and becomes pronounced by $t \approx 8.4 - 8.5s$ , after which the full-system response departs rapidly. This inside/outside split is the qualitative signature expected for a subcritical Hopf scenario with an unstable separating cycle. At the same time, the comparison confirms that the actual escape threshold of the full system is not identical to the local center-manifold cycle radius, but remains consistently organized by it as a local indicator.

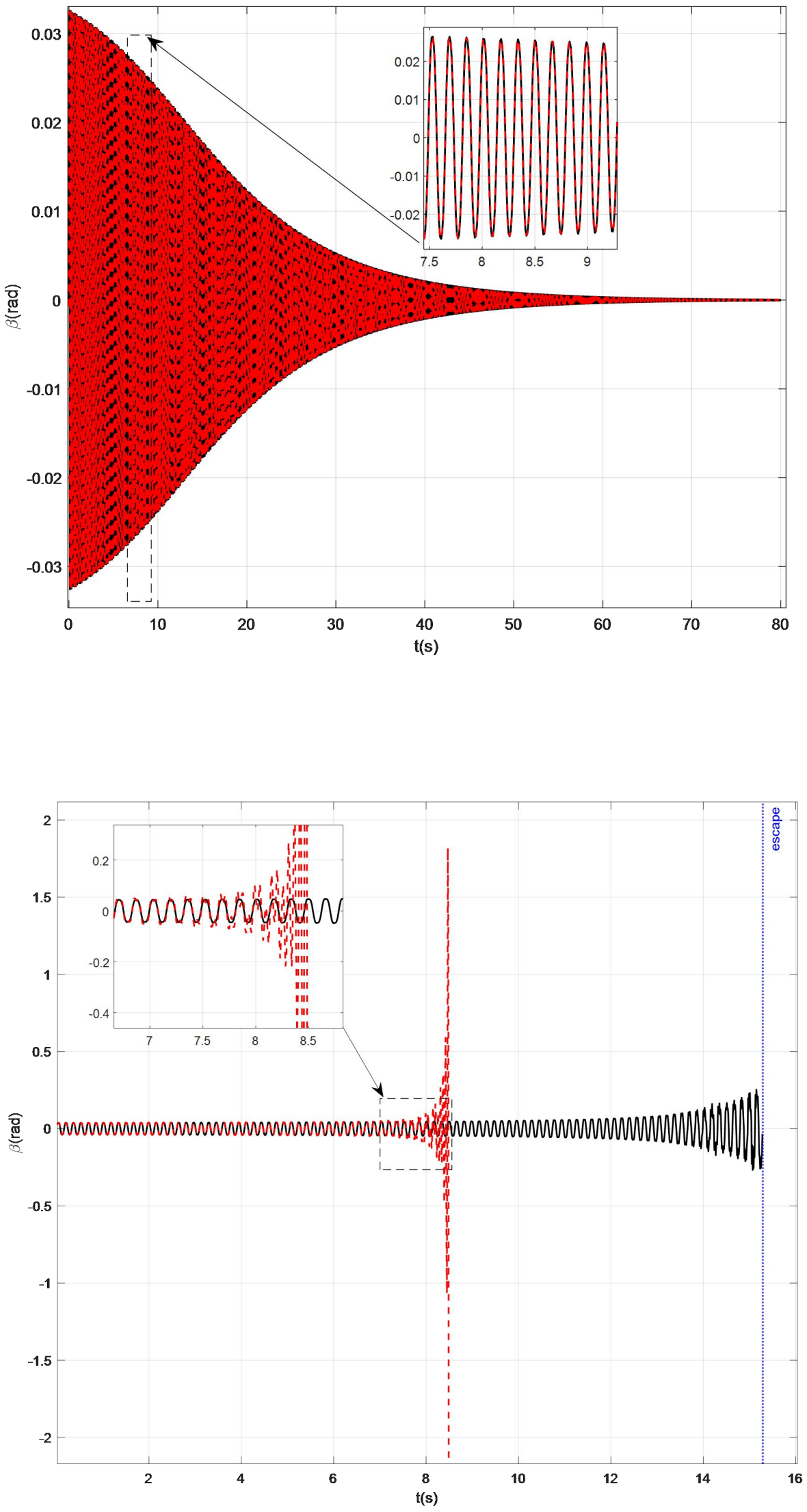

**Fig. 3** Time-domain responses for initial conditions chosen inside (upper) and outside (lower) the predicted subcritical-cycle threshold. Insets show the highlighted local windows used to assess agreement and onset of departure.

The inside initial condition remains attracted to the equilibrium, whereas the outside initial condition departs from the local neighborhood and eventually escapes. This inside/outside split is the qualitative signature expected from a subcritical Hopf scenario with an unstable separating cycle. In the present example, the transition occurs near $1.2r_{\mathrm{RG}}$ along the center direction, which is consistent with the reduced equation as a local indicator, while also illustrating that the actual full-system escape threshold is not identical to the local center-manifold cycle radius.

To further confirm the existence of the unstable cycle itself, A periodic orbit of the full collocation system was computed using the RG prediction as an initial guess. The corrected periodic solution is shown in Fig. 4, together with the corresponding RG-predicted cycle in the $(\beta, \dot{\beta})$ plane. The collocation result confirms that a nontrivial periodic orbit indeed exists in the neighborhood predicted by the RG reduction. In the reduced local picture, this orbit is unstable in the radial center-manifold direction and attracting in the transverse stable directions, which is the expected geometry of the subcritical cycle near a simple Hopf point.

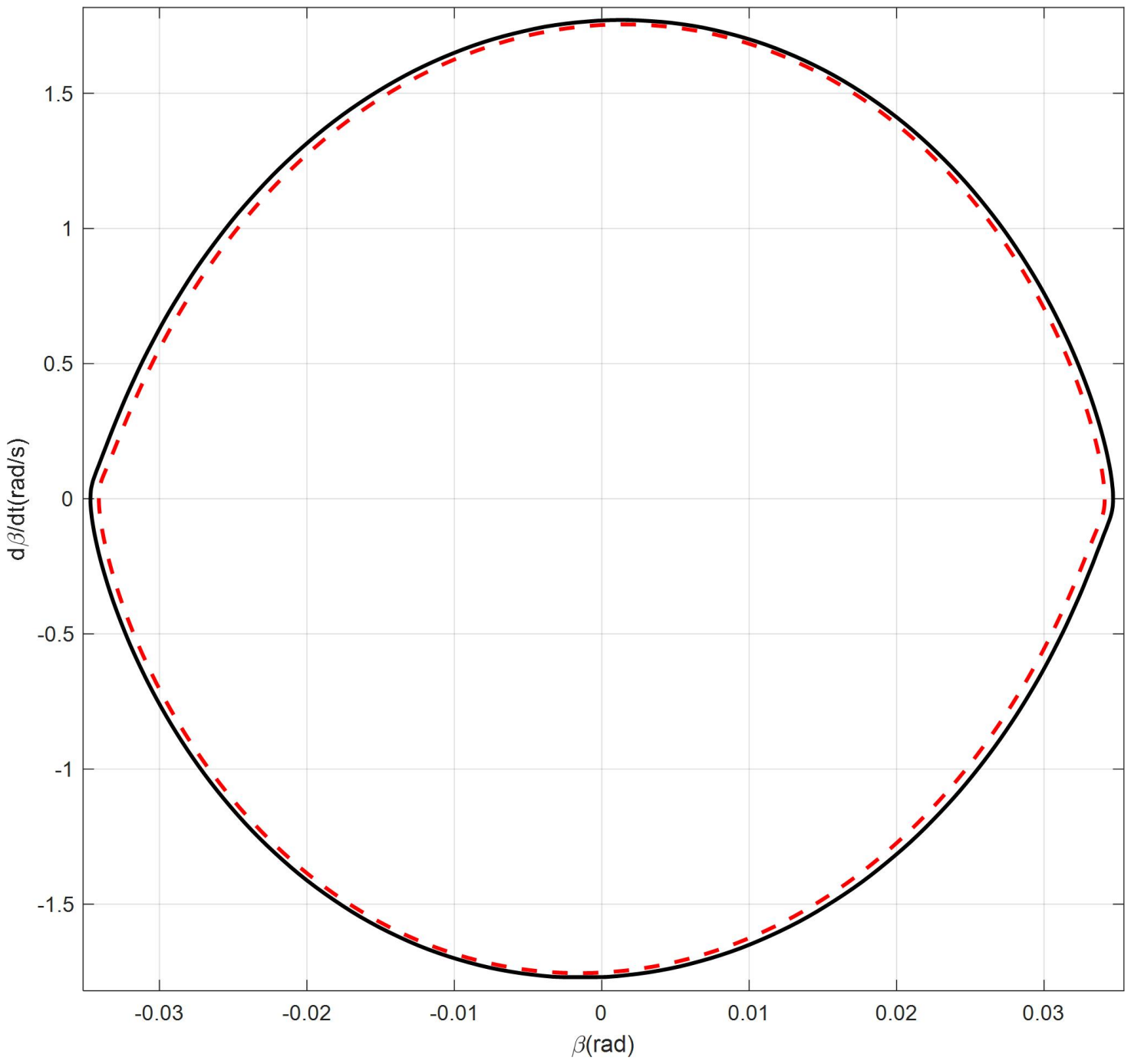

**Fig. 4** Comparison of the limit cycle in the $(\beta, \dot{\beta})$ plane between the RG prediction (red dashed) and the collocation-corrected periodic orbit (black solid).

The quantitative comparison is summarized in Table 2. The RG-predicted radius is $1.1349$, while the collocation-corrected cycle has period $0.16265$, compared with the RG estimate $0.16288$, corresponding to a relative period error of about $1.41\times10^{-3}$. The $\beta$ -amplitude changes from 0.0341 in the RG prediction to 0.0346 in the collocation-corrected cycle, giving a relative amplitude error of $1.645\times10^{-2}$. The $\dot{\beta}$ -amplitude changes from $1.7552$ to $1.7709$, corresponding to a relative error of $8.83\times10^{-3}$. The phase discrepancy also remains small. For the present purpose, this level of agreement is sufficient: the reduced equation not only classifies the local bifurcation correctly, but also locates the nearby unstable cycle with useful local accuracy.

A second validation concerns the time-domain discrepancy between the RG reconstruction and the direct simulation of the full nonlinear system. Instead of plotting the reconstructed envelope itself, Fig. 5 reports the corresponding absolute-error and relative-error histories, where the magenta curve denotes the absolute error and the cyan curve denotes the relative error. For the inside case, both error measures remain uniformly small over the local time window, which is consistent with the close agreement already visible in the inset of Fig. 3. For the outside case, the two responses remain close in the early stage, but the discrepancy begins to become visibly nonnegligible at about $t\approx8.2s$, and then grows rapidly once the trajectory exits the local neighborhood around $t\approx8.4-8.5s$. The corresponding trajectory-level error statistics are summarized in Table 3.

**Table 2** Comparison between the RG-predicted cycle and the collocation-corrected periodic orbit

| RG radius | $T_{RG}$ | $T_{full}$ |
|---|---|---|
| 1.1349 | 0.16288 | 0.16265 |
| $\beta$-amplitude RG | $\beta$ -amplitude full | Relative $\beta$ -amplitude error |
| 0.0341 | 0.0346 | $1.645\times10^{-2}$ |
| $\dot{\beta}$ -amplitude RG | $\dot{\beta}$ -amplitude full | Relative $\dot{\beta}$ -amplitude error |
| 1.7552 | 1.7709 | $8.83\times10^{-3}$ |

A second verification concerns the reconstructed time-domain dynamics. Fig. 5 compares the envelope evolution predicted by the RG reconstruction with the direct numerical simulation of the full nonlinear system. For the inside initial condition, the reduced and full responses remain nearly indistinguishable on the local time window, with a relative RMS signal error of $1.37\times10^{-3}$. For the outside initial condition, the error naturally grows once the trajectory leaves the local neighborhood; nevertheless, the reduced model still captures the correct qualitative trend, namely that the response departs once the initial amplitude exceeds the unstable threshold. The trajectory summary is reported in Table 3.

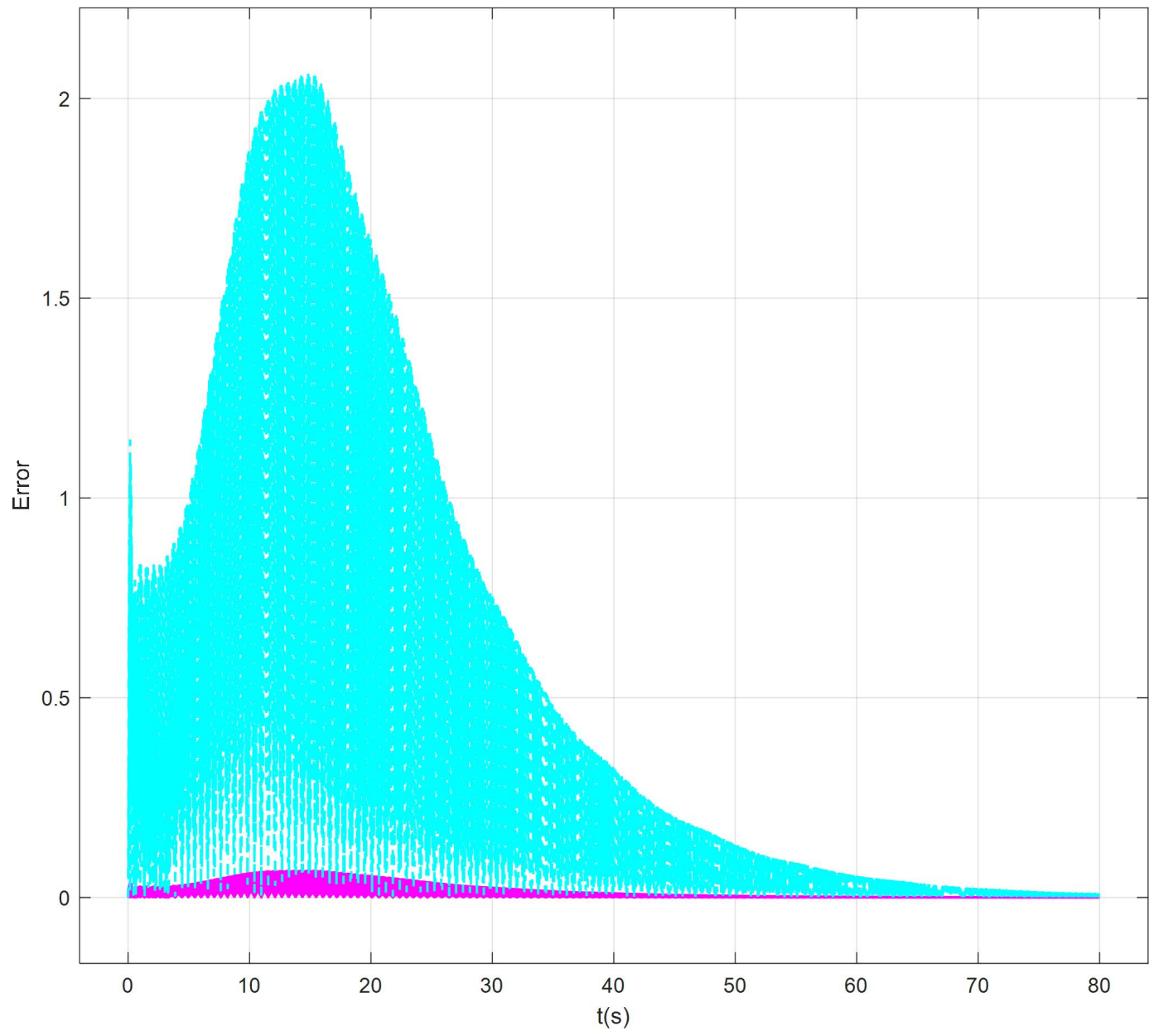

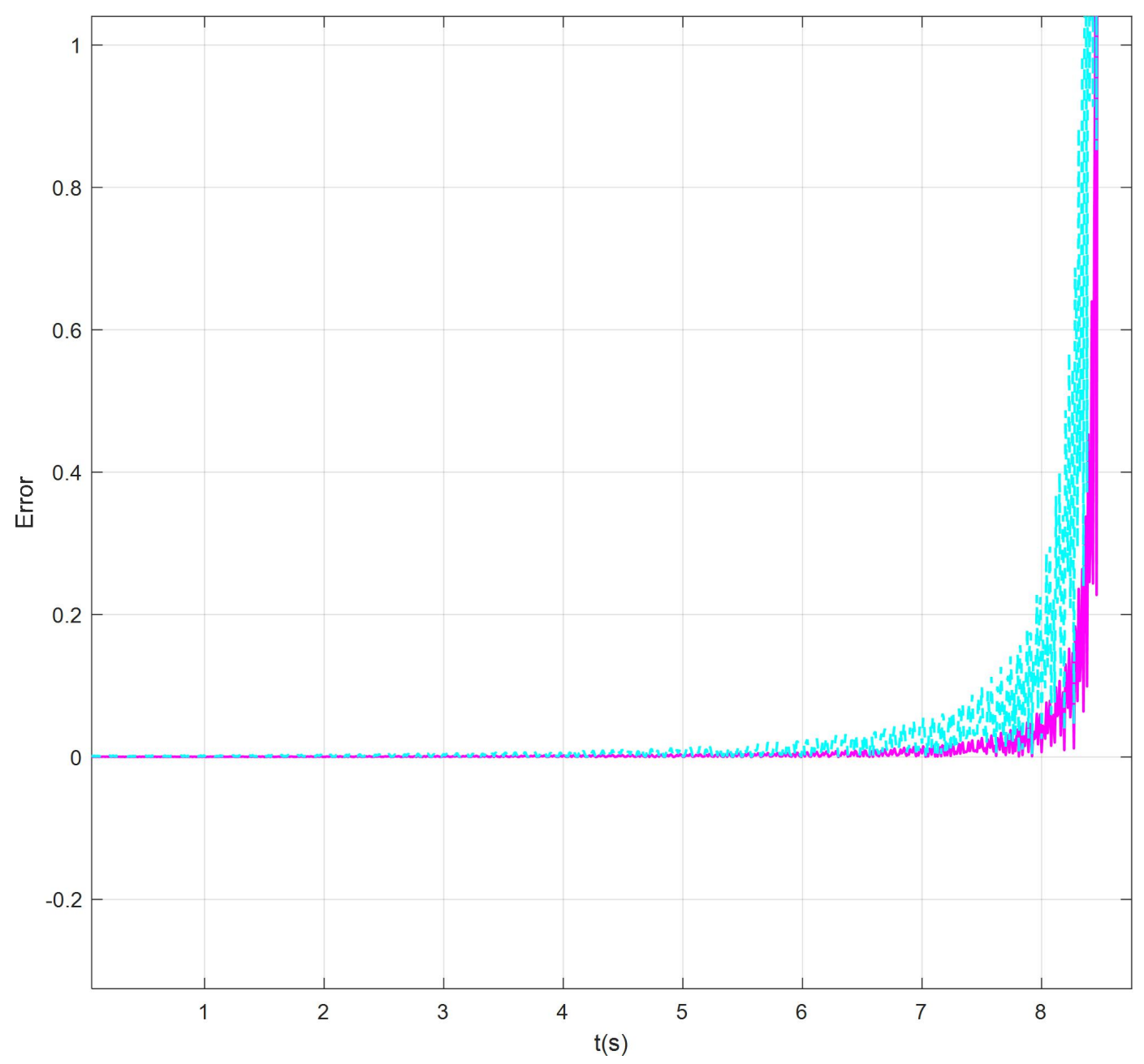

**Fig. 5** Relative-error (cyan) and absolute-error (magenta) histories between the RG reconstruction and the full-system simulation for the inside/outside initial conditions. Upper: inside case; lower: outside case.

**Table 3** Trajectory-level error summary for the RG reconstruction relative to the full-system response

| Case | Initial radius | Escaped | Escape time | RMS error | Max error | Relative RMS error |
|---|---|---|---|---|---|---|
| inside | 1.0781 | false | – | $2.1034\times10^{-5}$ | $6.7146\times10^{-5}$ | $6.458\times10^{-4}$ |
| outside | 1.3051 | true | 15.288 | $1.2489\times10^{-1}$ | 2.3138 | $4.6781\times10^{-1}$ |

Taken together, these results support the intended use of the RG equation as a local Hopf/LCO descriptor. In this example it successfully predicts the subcritical character of the bifurcation, provides a practically useful estimate of the unstable finite amplitude threshold on the center manifold, and reconstructs the nearby transient amplitude evolution with good local accuracy.

### 4.2.2 structural-mode substitution versus the true center eigenspace

The discussion next turns to a question that is directly relevant to reduced-order aeroelastic modeling: whether an undamped structural modal basis can replace the true center eigenspace when the objective is local Hopf sensitivity analysis. The comparison made here is not fully equivalent to a quasi-steady elimination of aerodynamic states, and therefore should not be interpreted as a general statement about all engineering ROM practices. Rather, it is intended as a focused diagnostic test for the present application.

The quantity of interest is the linear coefficient of the RG equation associated with the stiffness perturbation $k_\beta$. As noted above, this coefficient is the projected sensitivity

$$W_c^* \frac{\partial A}{\partial k_\beta} V_c$$

whose real part gives the leading drift of the critical growth rate and whose imaginary part gives the leading drift of the critical frequency. In the present example, this quantity is extracted directly from the RG equation as

$$1.78645 + 0.759461\, i$$

To construct a structural surrogate, The true center mode is cropped to its structural coordinates and compared with the corresponding undamped structural mode. The modal similarity is high: for the representative point reported in Table 3 (b), the right- and left-mode similarities remain above 0.99 in MAC and cosine-based measures. At first sight, such high similarity might suggest that the structural mode should yield a comparable sensitivity prediction. However, the actual sensitivity comparison shows the opposite. As reported in Table 3 (a), the structural-mode-based estimate of the real part is

$$\Re\left(\lambda'_{\text{struct}}\right) = -3.37\times10^{-4},$$

whereas the true cropped-center estimate is

$$\Re\left(\lambda'_{\text{true}}\right) = 1.7864.$$

Thus, the structural surrogate not only incurs a large quantitative error, but even predicts the wrong sign of the real-part sensitivity. The imaginary part is also substantially underestimated. The contrast is particularly striking because it occurs despite the high similarity of modal shapes.

**Table 4** Single-point comparison of eigenvalue sensitivity obtained from the structural-mode surrogate and the true cropped center modes

**(a)**

| | Re | Im | Relative error (real) | Relative error (imag) |
|---|---|---|---|---|
| struct | -0.00033709 | 0.074381 | 1.0002 | 0.90206 |
| true | 1.7864 | 0.75946 | | |

(b)

| $MAC_V$ | $MAC_W$ | $\mathrm{COS}_V$ | $\mathrm{COS}_W$ |
|---|---|---|---|
| 0.99209 | 0.99002 | 0.99604 | 0.995 |

The reason becomes clearer when the sensitivity is decomposed into channel-wise contributions. The component and channel breakdowns are listed in Table 5. In the structural-mode calculation, the real-part contribution from the stiffness channel is almost negligible, while the damping-related part contributes a small negative real drift, leading overall to a nearly zero or slightly negative sensitivity. By contrast, the true center-based projection retains a strong positive destabilizing contribution in the critical projection. In other words, the parameter perturbation is not felt merely through the geometric resemblance of the mode shape, but through how the perturbation is weighted by the full right-left critical pair of the coupled aeroelastic operator.

**Table 5** Channel-wise decomposition of the real-part sensitivity for the structural-mode surrogate and the true center-based projection

| | Stiff | Damp |
|---|---|---|
| Struct Re | 0 | -0.00033709 |
| True Re | 1.7912 | -0.0047 |

This distinction is important in practice. It does not mean that structural-mode ROMs are useless for response reconstruction; on the contrary, they may still be effective for many simulation tasks. What the present example shows is more specific: for local Hopf sensitivity analysis, modal-shape similarity alone is not sufficient. Even when the cropped critical mode and the structural mode appear almost identical, the sensitivity of the critical eigenvalue can still be qualitatively mis-predicted if the true adjoint projection of the coupled aeroelastic operator is not respected.

To assess whether this is merely an isolated pointwise pathology, Both Hopf branches were scanned over the $\left(U, x_a\right)$ window. The resulting trends are shown in Fig. 6. Branch 1 corresponds to the control-surface-dominated flutter mode, and Branch 2 to the bending-dominated flutter mode.

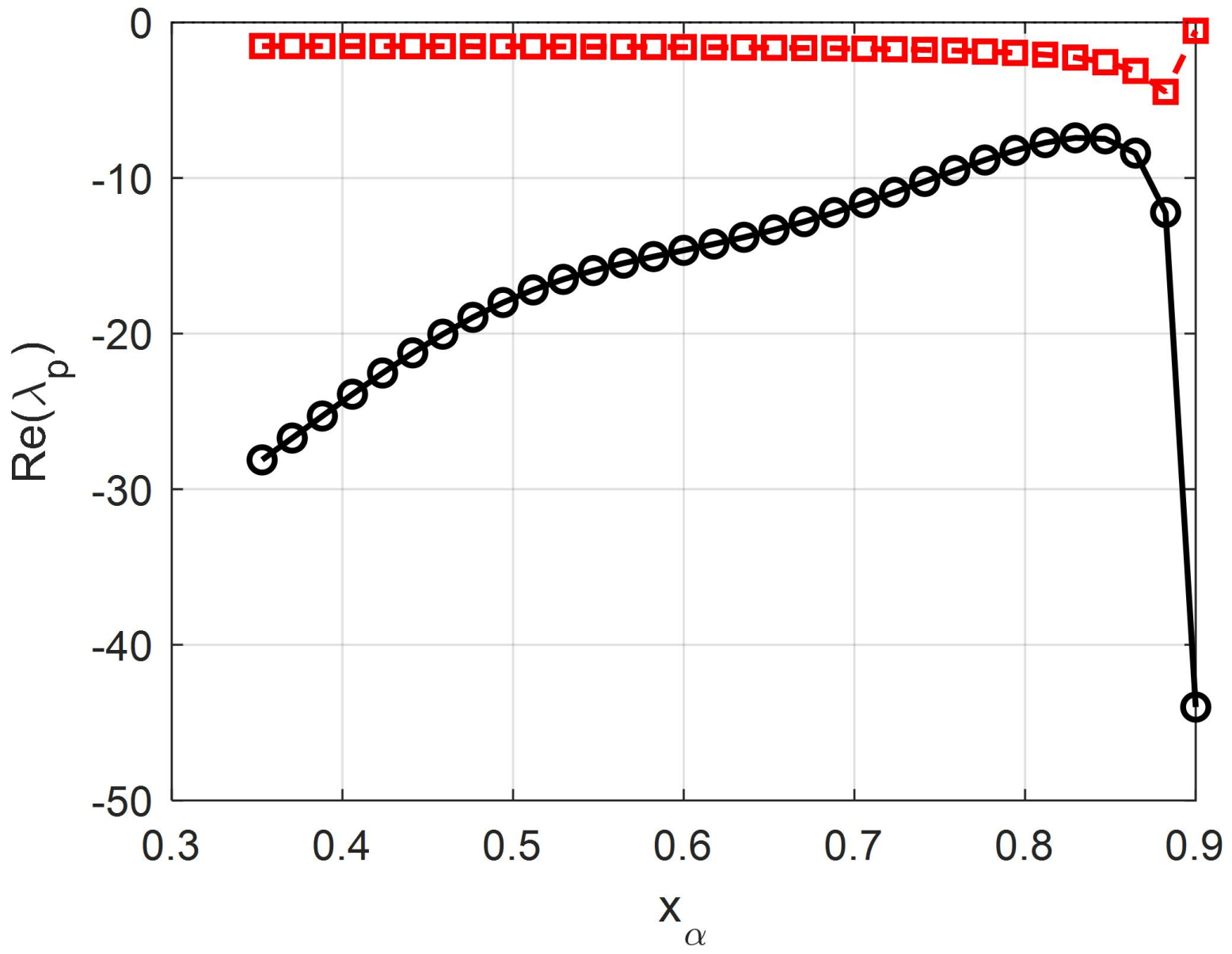

(a)

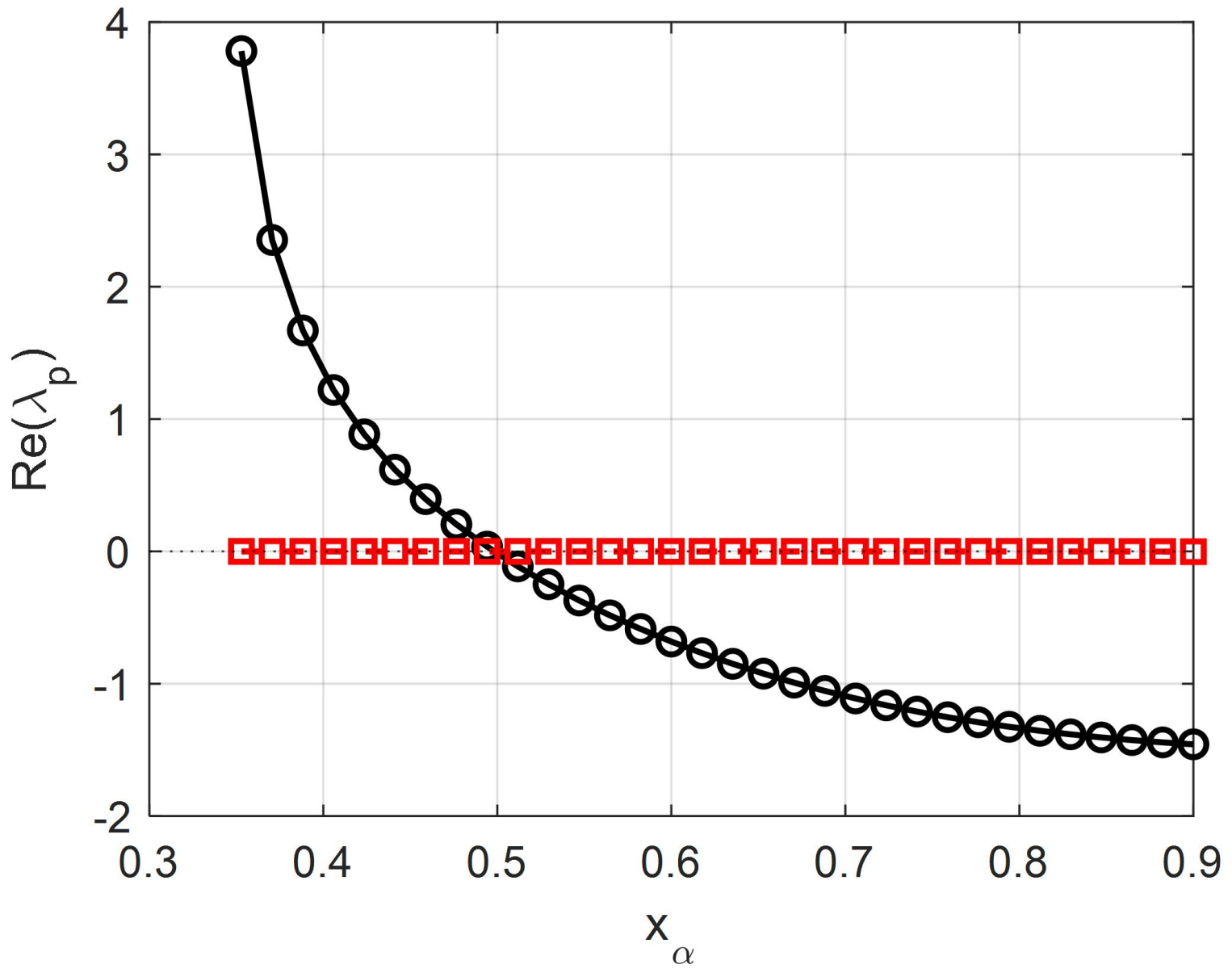

(b)

**Fig. 6** Sensitivity of the two Hopf branches to $k_\beta$ : true center-eigenspace projection versus

structural-mode surrogate. (a) Branch 1. (b) Branch 2. Red: structural-mode surrogate; black: true center-based result.

On both branches, the structural surrogate produces nearly flat sensitivity curves, whereas the true center-based sensitivities vary substantially with $x_a$; for the bending branch, the real-part sensitivity even changes sign over the scanned window. Hence the discrepancy is not an accidental numerical mismatch at a single point, but a persistent limitation of structural-mode substitution when the target quantity is the local bifurcation sensitivity.

#### 4.2.3 branch-dependent effects of nonlinear stiffness on the Hopf cubic coefficient

Finally, the manner in which different nonlinear stiffness mechanisms modify the local post-bifurcation tendency along the two Hopf branches is examined. In the present computations, the direct cubic structural stiffnesses are set to

$$k_{h3} = 100, \qquad k_{\alpha 3} = 25, \qquad k_{\beta 3} = 45,$$

and the corresponding contributions to the RG Hopf cubic coefficient are evaluated branchwise. Because the cubic term in the RG amplitude equation depends linearly on the cubic nonlinear tensor, the reduced Hopf cubic coefficient can be decomposed additively into the parts induced by the plunge, pitch, and control-surface cubic stiffnesses. This decomposition makes it possible to compare, for each branch, which physical stiffness mechanism is actually active in the local Hopf dynamics.

Fig. 7 shows these branchwise cubic-stiffness contributions. The two branches display clearly different patterns, which is consistent with their different modal character. see Fig. 7(a), Branch~1 is the control-surface-dominated flutter branch. Along this branch, the dominant contribution is the one associated with the control-surface cubic stiffness $k_{\beta 3}$ : the corresponding curve remains negative throughout the scanned interval, with a relatively large magnitude on the left side of the branch and a gradual recovery toward zero as $x_\alpha$ increases. A mild turning tendency appears near the right end, but the contribution remains negative. By contrast, the pitch-related contribution associated with $k_{\alpha 3}$ is much smaller in magnitude. The zoomed view shows that it is positive for smaller $x_\alpha$, decreases steadily as $x_\alpha$ increases, and becomes very close to zero around the upper part of the interval before turning slightly negative near the right endpoint. The plunge-related contribution associated with $k_{h3}$ stays essentially at zero over the whole branch and is therefore negligible at the present scale. Thus, for Branch~1, the leading cubic correction is governed primarily by the control-surface stiffness, while the pitch stiffness provides only a weak secondary correction and the plunge stiffness is practically inactive.

Branch~2 shows a different balance, see Fig. 7(b). This branch is bending-dominated, and all three cubic stiffnesses leave visible signatures in the reduced coefficient, although with clearly different importance. The largest contribution is now the pitch-cubic part associated with $k_{\alpha 3}$, which remains negative over the entire scanned interval and exhibits a broad minimum in the middle portion of the branch before recovering slightly for larger $x_\alpha$. The plunge-cubic contribution associated with $k_{h3}$ is smaller but still non-negligible: it is positive over the left part of the branch and decays gradually toward zero as $x_\alpha$ increases. The control-surface-cubic contribution associated with $k_{\beta 3}$ is weaker than the pitch contribution but is not negligible; the enlarged view shows that it is positive near the left side of the interval, decreases toward zero, and changes sign in the upper part of the branch, after which it remains slightly negative. Therefore, on Branch~2, the local cubic tendency reflects a competition among several mechanisms rather than the dominance of a single one: pitch cubic stiffness provides the main negative contribution, plunge cubic stiffness contributes a weaker positive part over the left portion of the branch, and control-surface cubic stiffness changes role along the branch.

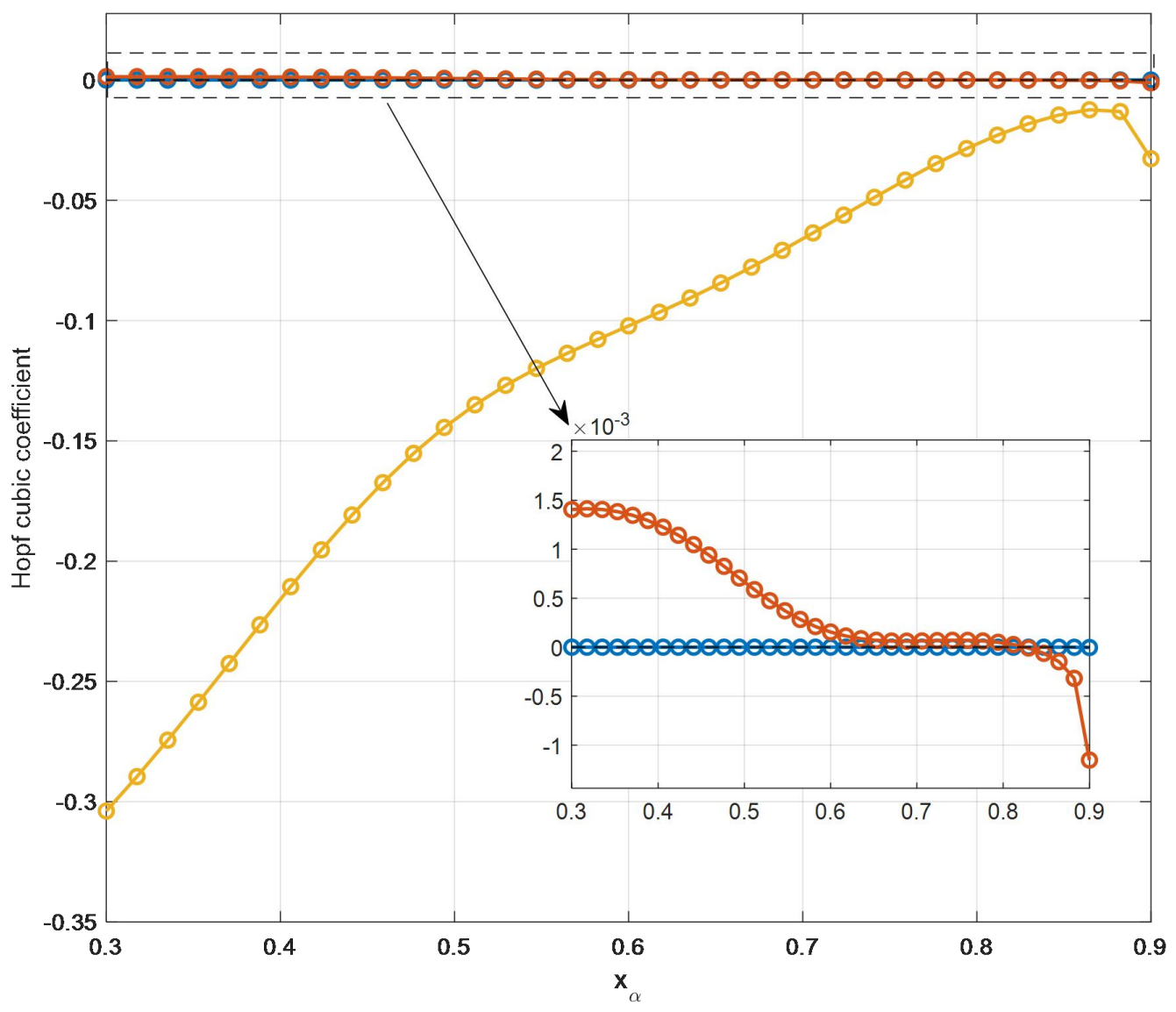

(a)

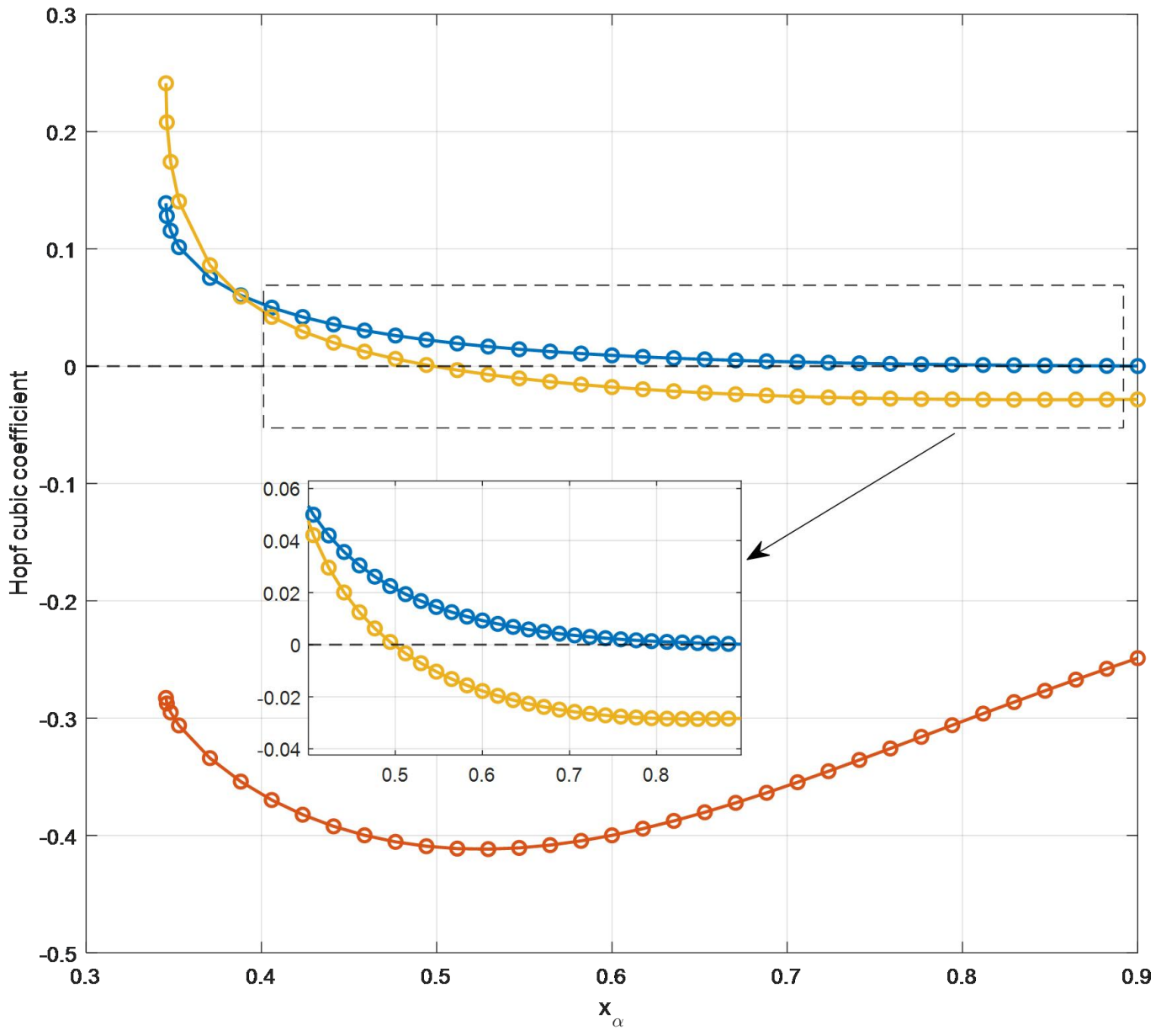

(b)

**Fig. 7** Branchwise decomposition of the RG Hopf cubic coefficient: blue $k_{h3}$, red $k_{\alpha 3}$, yellow $k_{\beta 3}$; (a) Branch 1, (b) Branch 2; dashed, zero.

Taken together, Fig. 7 confirms that the Hopf cubic coefficient is both branch-dependent and parameter-dependent. The same physical nonlinearity may dominate one flutter branch but become secondary or even sign-changing on another. In this sense, the reduced cubic coefficient is not determined by the restoring law alone; it is determined by how that restoring law is projected onto the critical eigenspace associated with a specific branch.

To further examine the role of noncritical-mode mediation, A quadratic structural coupling term of $h\alpha$ -type is additionally introduced. More precisely, we supplement the restoring force by a conservative quadratic potential that generates a coupling coefficient denoted here by $k_{h\alpha}$ ,which has the form:

$$V_{ha} = \frac{1}{10b}\frac{\sqrt{K_h K_\alpha}}{2} k_{ha} h^2 \alpha, K_h = m_w \omega_h^2, K_\alpha = m_w \omega_\alpha^2 r_\alpha^2 b_\alpha^2$$

This term is introduced on purpose, not because it is the main nonlinearity in the present airfoil model, but because it provides a clean probe of a mechanism that direct cubic terms do not reveal: a quadratic nonlinearity can affect the Hopf cubic coefficient only through lower-order excitation of noncritical directions followed by feedback onto the critical oscillation. In that sense, $k_{h\alpha}$ is used here as a diagnostic nonlinearity for assessing the influence of stable-mode mediation in the local reduction. Such quadratic couplings are also not entirely artificial from an engineering viewpoint, since mixed geometric or offset-induced restoring effects may generate analogous cross terms in more detailed structural models. In the present test, The coefficient is set to

$$k_{h\alpha} = 1,$$

and examine the induced contribution to the Hopf cubic coefficient under progressive modal enrichment.

The results are shown in Fig. 8. The blue curve corresponds to the center pair only, while the red, yellow, purple, and green curves denote reductions enriched by one, two, three, and four additional stable/unstable modal pairs, respectively. The enrichment study shows that this induced contribution is the one most sensitive to noncritical-mode mediation, which is precisely why it is useful here.

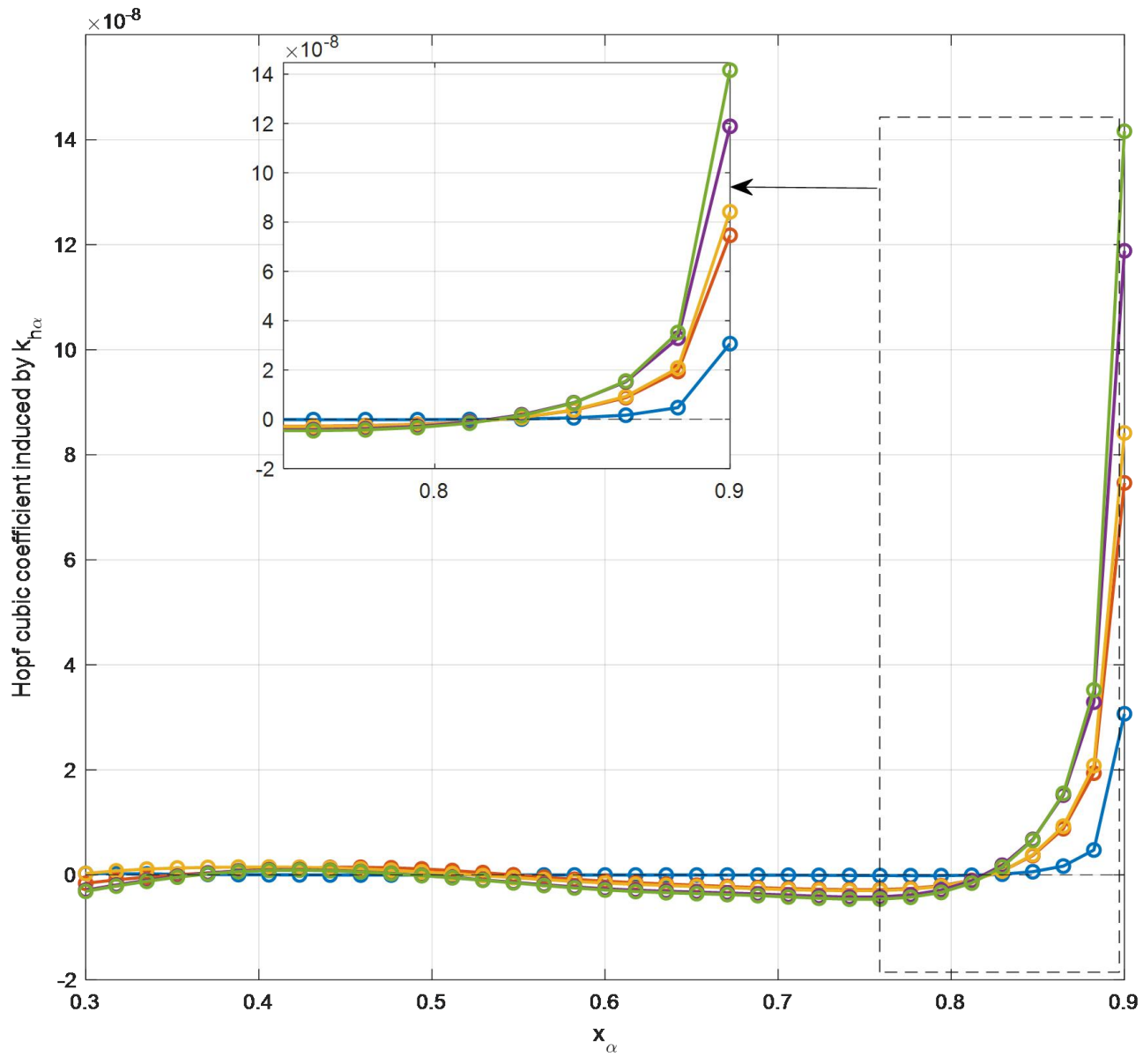

(a)

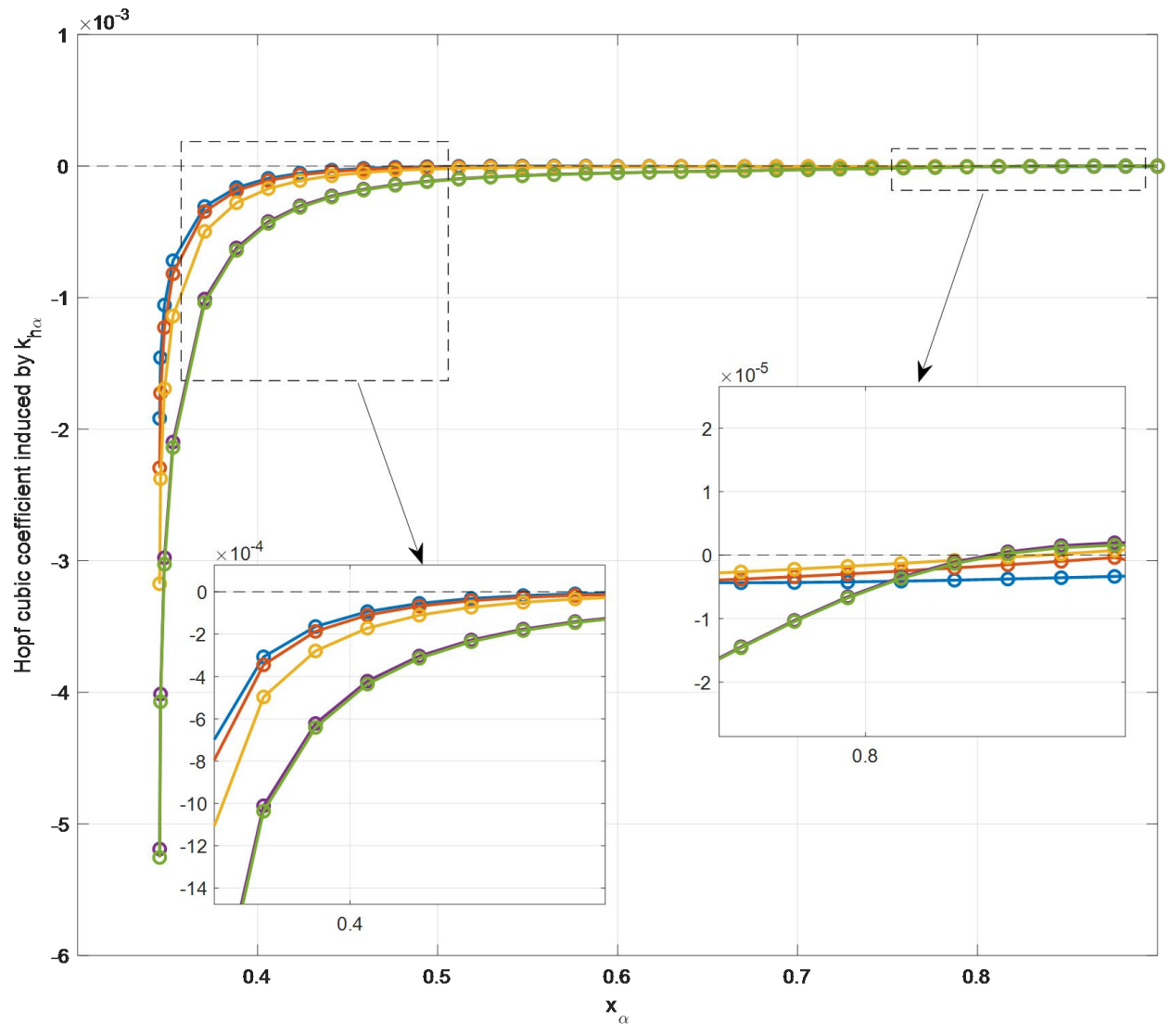

(b)

**Fig. 8** contribution of the $k_{h\alpha}$ -type quadratic coupling to the RG Hopf cubic coefficient with mode enrichment. (a) Branch 1. (b)Branch 2. blue, center pair only; red/yellow/purple/green, plus 1/2/3/4 modal pairs.

For Branch 1, Fig. 8(a) show that the induced contribution stays very small over most of the branch, and the center-only model already captures the fact that its effect is weak in the main part of the parameter interval. However, the enlarged view near the right end shows a sharp positive growth as $x_\alpha$ approaches the branch endpoint. In that localized region, the enriched reductions separate clearly from the center-only curve, and the predicted contribution increases systematically as more noncritical modal pairs are retained. At the same time, the difference between the three-pair and four-pair enrichments is already small compared with the gap between the center-only and enriched curves, which indicates an emerging saturation of the enrichment effect. Hence, on Branch 1, $k_{h\alpha}$ does not modify the cubic coefficient significantly over most of the branch, but its induced effect can be amplified near the right endpoint by mediation through noncritical modes.

For Branch 2, Fig. 8(b) the induced contribution is negative over most of the scanned interval, with the strongest magnitude near the left side of the branch and a gradual relaxation toward zero as $x_\alpha$ increases. In the left portion of the branch, the separation among the enriched curves is clear, showing that noncritical modes materially deepen the induced negative contribution in the region where the coupling is strongest. Farther downstream, all curves move toward zero and become very close to one another. The enlarged view near the upper part of the branch shows that the induced contribution is no longer strictly negative there: it crosses through zero around $x_\alpha \approx 0.86$ and becomes slightly positive for the more enriched reductions, whereas the center-only result remains slightly negative at essentially the same location. The magnitude in this region is small, of order $10^{-6}$, so this sign change should not be over-interpreted. Still, because the effect of $k_{h\alpha}$ enters through a quadratic mechanism, its contribution scales with the square of the associated coefficient. Therefore, although the present test uses $k_{h\alpha}=1$ only as a diagnostic level, the induced effect need not be negligible a priori if a comparable coefficient scale were realized in a model where such coupling is physically relevant.

These results help separate two qualitatively different mechanisms. The contributions from $k_{h3}$, $k_{\alpha 3}$, and $k_{\beta 3}$ are direct cubic effects: they are already visible at the level of the center reduction, and their branch dependence can be interpreted through direct projection onto the critical eigenspace. By contrast, the $k_{h\alpha}$ contribution is an induced cubic effect: it is generated indirectly through quadratic excitation of noncritical directions and their subsequent slaving back into the reduced cubic coefficient. The mode-enrichment comparison therefore shows that retained noncritical modes are not only relevant for state reconstruction; in some parameter regions they can also modify the Hopf descriptor itself.

Overall, the present branchwise results support two conclusions. First, the effect of nonlinear stiffness on the Hopf cubic coefficient is strongly branch-specific: a mechanism that dominates one branch may be weak, competing, or even sign-changing on another. Second, when the nonlinear mechanism enters through quadratic coupling such as $k_{h\alpha}$, reliable local Hopf/LCO descriptors may require explicit noncritical-mode mediation, even if the induced effect remains comparatively small in absolute magnitude over most of the branch.

A related observation helps explain why the black curves in Fig.6 and the yellow curves in Fig.7 display similar branchwise trends. In the present model, both the real-part sensitivity with respect to the control-surface linear stiffness $k_{\beta 0}$ and the direct $k_{\beta 3}$ contribution to the Hopf cubic coefficient are governed by the same control-surface displacement channel in the first-order system. From Appendix C, the linear coefficient of the reduced Hopf equation is $\sigma = W^* A_\mu V$, where $\mu$ denotes the unfolding parameter associated here with $k_{\beta 0}$. Using the first-order form in Appendix D

$$A_\mu = \begin{bmatrix} 0 & 0 \\ -M^{-1}K_\mu & -M^{-1}C_\mu \end{bmatrix}.$$

If one isolates the stiffness part relevant to the present comparison, then differentiating the structural stiffness matrix with respect to $k_{\beta 0}$ gives a perturbation acting only in the control-surface restoring channel, so that $K_\mu V_s \propto e_\beta V_3$, where $V_s$ is the structural-displacement block of $V$, $V_3$ is its control-surface displacement component, and $e_\beta$ is the unit vector of that channel. Hence the stiffness-induced part of the sensitivity has the form $W^* A_\mu V \sim -W_v^* M^{-1} e_\beta V_3$, up to fixed coefficient factors. Now consider the direct cubic restoring force generated by $k_{\beta 3}$. In the first-order system, this contribution enters only through the same control-surface channel

$$F_{\beta 3}(x) = \begin{bmatrix} 0 \\ -M^{-1}\left(e_\beta c_\beta \beta^3\right) \end{bmatrix}$$

with $c_\beta$ collecting constant prefactors. Evaluating it on the leading critical oscillation $x \approx VC + \bar{V}\bar{C}$ gives $\beta \approx V_3 C + \bar{V}_3 \bar{C}$, and therefore $\beta^3 = V_3^3 C^3 + 3V_3^2 \bar{V}_3 C^2 \bar{C} + 3V_3 \bar{V}_3^2 C \bar{C}^2 + \bar{V}_3^3 \bar{C}^3$. The resonant term entering the reduced Hopf equation is the $C^2\bar{C}$ term, namely

$$3V_3^2 \bar{V}_3 = 3V_3 \,|\, V_3 \,|^2 .$$

After projection by $W^*$, the direct $k_{\beta 3}$ contribution therefore has the structure $\gamma_{\beta 3}^{\text{dir}} \sim -W_v^* M^{-1} e_\beta \left(3V_3 \,|\, V_3 \,|^2\right)$, again up to fixed coefficient factors. Therefore, apart from the additional positive weighting $3\,|\, V_3 \,|^2$, the direct $k_{\beta 3}$ contribution is filtered by the same control-surface channel and the same critical right-left modal factors as the stiffness sensitivity with respect to $k_{\beta 0}$. This explains why the black curves in Fig. 6 and the yellow curves in Fig. 7 show similar signs, turning tendencies, and, on the bending-dominated branch, nearly the same zero crossing near $x_\alpha \approx 0.49$.

### 4.2.4 Summary of the numerical findings

The numerical study supports three conclusions relevant to local nonlinear aeroelastic analysis near flutter. First, the RG reduction provides a practically useful local description of a subcritical Hopf bifurcation, including the existence and approximate location of a nearby unstable limit cycle and the corresponding inside/outside transient behavior. Second, the linear coefficient of the reduced equation preserves the critical eigenvalue sensitivity information of the full coupled system, whereas replacing the true center eigenspace by a structural modal basis may lead to qualitatively incorrect sensitivity predictions even when the modal shapes appear highly similar. Third, the reduced cubic coefficient offers a compact way to disentangle how different nonlinear stiffness mechanisms alter the local post-bifurcation tendency, while also making clear when stable-mode mediation must be retained explicitly.

## 5. Conclusion

This work developed a renormalization-group-based local reduction framework for Hopf bifurcation and near-onset limit-cycle oscillation analysis in nonlinear aeroelastic systems. The formulation is organized for polynomial/tensor nonlinearities and yields, in a direct algorithmic form, a reduced complex amplitude equation together with the coefficients governing the Hopf threshold, local criticality, and leading-order amplitude/frequency trends. A companion slow-manifold approximation was also introduced, allowing selected stable directions to be retained as static coordinates when a more informative reconstruction is desired.

A first contribution of the paper is therefore methodological. Appendix A recasts the RG construction as an explicit reduction procedure suitable for local Hopf analysis, rather than only as a formal perturbation calculation. In this sense, the method may be viewed as a particular coordinate/normal-form choice on a local invariant manifold, but one that remains convenient for direct coefficient extraction and implementation with polynomial nonlinear terms.

A second contribution is theoretical, but it should be understood in a controlled and problem-oriented sense. Building on earlier RG manifold results, Appendix B provides a Hopf-oriented justification for the truncated RG reduction used here. Under standard assumptions on spectral separation, normal hyperbolicity, and Hopf nondegeneracy, the RG map defines an approximate invariant manifold whose reduced flow is $C^r$ -close to that on a true local invariant manifold. This implies that the Hopf point and its supercritical/subcritical character predicted by the truncated RG system persist in the full system, with parameter and coefficient errors of the same asymptotic order as the truncation. The point here is not to replace classical center-manifold and normal-form theory, but to clarify why the RG-based reduction used in the paper is mathematically consistent for the present Hopf setting.

The numerical results then illustrate what this local description can reveal in practice. First, in the subcritical example, the RG reduction gives a useful local prediction of the nearby unstable cycle and of the associated inside/outside transient behavior. Second, the study of the linear RG coefficient shows that local Hopf sensitivity is controlled by the true right--left critical projection of the coupled aeroelastic operator; an undamped structural modal surrogate may therefore give not only a large quantitative error but even an incorrect sign in the predicted sensitivity. Third, the branchwise decomposition of the Hopf cubic coefficient shows that the effect of nonlinear stiffness is strongly branch-dependent: a cubic stiffness mechanism that is dominant on one flutter branch may be weak or sign-changing on another. In addition, the calculations with the quadratic coupling term show that some induced cubic effects are mediated by noncritical modes and may therefore be missed by an overly restrictive center-only reduction.

A further useful observation is that, in the present model, the sensitivity with respect to the control-surface linear stiffness and the direct contribution of the control-surface cubic stiffness to the Hopf cubic coefficient exhibit similar branchwise trends because both are filtered through the same control-surface displacement channel and the same critical right--left modal factors. This does not imply an exact identity between the two quantities, but it provides a concise interpretation of the similar signs, turning tendencies, and zero-crossing behavior observed in the numerical results.

Overall, the present work suggests that the RG coefficients are not merely formal perturbation quantities. When used with appropriate care, they provide compact local descriptors of flutter onset and early post-critical behavior, and they help separate direct critical-mode effects from contributions mediated by noncritical directions. Future work will focus on extending the same framework to richer aerodynamic nonlinearities, higher-order reductions, and larger discretized models, including finite-element-based implementations.

**Funding**

The authors declare that no funds, grants, or other support were received during the preparation of this manuscript.

**Competing Interests**

The authors declare that they have no relevant financial or non-financial interests to disclose.

## Author Contributions

Gelin Chen: conceived the study, performed theoretical analysis and numerical design, wrote the manuscript, and prepared all figures and text. Chao Yang and Chen Song: contributed to discussions, advised on methodology, and provided critical revisions of the manuscript. All authors read and approved the final manuscript.

## Data Availability

The datasets generated and/or analyzed during the current study are available from the corresponding author on reasonable request.

## Appendix

### Appendix A. RG Algorithm for Polynomial Nonlinearities

#### A.1 Problem Statement and Perturbation Hierarchy

Consider the weakly nonlinear system

$$\dot{x} = Ax + \varepsilon g(x) \tag{A.1}$$

where $x \in \mathbb{R}^n, 0 < \varepsilon \ll 1$, and $A$ is a constant real matrix. The nonlinearity $g$ is assumed to be polynomial in $x$; its precise structure will be specified below. We introduce the standard perturbation expansion

$$x(t) = x_0(t) + \varepsilon x_1(t) + \varepsilon^2 x_2(t) + \cdots \tag{A.2}$$

and substitute it into (A.1). By collecting terms at each power of $\varepsilon$ we obtain the perturbation hierarchy

$$\dot{x}_0 = Ax_0,$$
$$\dot{x}_i = Ax_i + G_i\left(x_0,\ldots,x_{i-1}\right), \quad i \geq 1 \tag{A.3}$$

where the forcing terms $G_i$ are determined by $g$ and the lower-order solutions. For completeness we list the first few $G_i$ explicitly, the general pattern is described in Appendix A.5. To make the multiple-scale bookkeeping transparent, we expand the nonlinearity as

$$g(x) = \varepsilon g_1(x) + \varepsilon^2 g_2(x) + \varepsilon^3 g_3(x) + \cdots \tag{A.4}$$

without imposing a one-to-one correspondence between the order in $\varepsilon$ and the polynomial degree in $x$. Each $g_k$ may collect several multilinear contributions of different degrees; this is convenient for engineering applications where the $\varepsilon$ -ordering is dictated by a chosen scaling rather than by degree alone.

Substituting (A.2)-(A.4) into (A.1) and using the chain rule yields, for the first few orders,

$$G_1 = g_1\left(x_0\right) \tag{A.5a}$$

$$G_2 = Dg_1\left(x_0\right)x_1 + g_2\left(x_0\right) \tag{A.5b}$$

$$G_3 = Dg_1\left(x_0\right)x_2 + \frac{1}{2}D^2 g_1\left(x_0\right)\left(x_1,x_1\right) + Dg_2\left(x_0\right)x_1 + g_3\left(x_0\right) \tag{A.5c}$$

and so on. Under the polynomial-tensor assumption, all derivatives of $g$ are again multilinear tensors; hence each $G_i$ can be assembled through a finite number of tensor contractions and additions.

In what follows we work in complex notation for convenience, but always impose the appropriate conjugacy conditions to guarantee that the physical solution $x\left(t\right)$ remains real.

**A.2 Spectral Assumptions**

We impose the following assumptions on the matrix $A$ and the nonlinearity $g$ : The spectrum of $A$ consists of a finite set of purely imaginary eigenvalues

$$\left\{\pm \mathrm{i}\omega_1, \ldots, \pm \mathrm{i}\omega_{N_c}\right\}, \tag{A.6}$$

and a finite set of stable eigenvalues with strictly negative real parts. There are no eigenvalues with positive real part; All eigenvalues of $A$ are semi-simple (diagonalizable). In particular, both the center and the stable spectra are free of Jordan blocks; Polynomial nonlinearity The function g is polynomial in $x$ (or, more generally, admits an absolutely convergent power series on the neighborhood of interest).

Let $U_c = \left[U_1, \ldots, U_{N_c}, \bar{U}_1, \ldots, \bar{U}_{N_c}\right]$ denote the matrix of right eigenvectors corresponding to the center eigenvalues, arranged in complex-conjugate pairs. Let $U_s$ collect the right eigenvectors corresponding to the stable eigenvalues. Similarly, let $V_c$ and $V_s$ denote the corresponding matrices of left eigenvectors. We scale them so that the bi-orthogonality relations

$$V_c^* U_c = I, \quad V_c^* U_s = 0, \quad V_s^* U_c = 0, \quad V_s^* U_s = I \tag{A.7}$$

hold. Here $(\cdot)^*$ denotes conjugate transpose.

Under these assumptions, the real subspace spanned by the center eigenvectors is invariant under the linear flow, and the restriction of $e^{At}$ to that subspace is a finite combination of pure rotations.

Remark. Assumption 3 (polynomial nonlinearity) is not essential for the RG method itself. the framework extends to smooth nonlinearities that are uniformly bounded on compact domains, and even to non-autonomous systems whose time-dependent forcing admits a Fourier expansion without accumulation points. We adopt the polynomial assumption in order to facilitate algorithmic implementation in finite-element-based applications; in practice it covers most engineering models of interest.

### A.3 Modal Representation

We introduce modal coordinates

$$y = V^{*}x, \quad x = Uy, \quad U = [U_c U_s], V = [V_c V_s], \tag{A.8}$$

so that the linear part becomes diagonal in blocks. Under our spectral assumptions, the solution to the leading-order equation

$$\dot{x}_0 = Ax_0$$

may be written as

$$x_0(t) = X(t)U_c C, \quad X(t) = e^{At}, \tag{A.9}$$

where $C \in \mathbb{C}^{2N_c}$ collects the complex amplitudes corresponding to the center eigenvectors. The amplitudes associated with negative-imaginary eigenvalues are treated as formally independent variables; the reality of $x_0$ is enforced by imposing the conjugacy constraints $C_{k+N_c} = \bar{C}_k$ for $k = 1, \ldots, N_c$.

Expressed in terms of the individual center eigenvectors, (A.9) reads

$$x_0(t) = \sum_{k=1}^{N_c} \left( C_k e^{\mathrm{i}\omega_k t} U_k + C_{k+N_c} e^{-\mathrm{i}\omega_k t} \bar{U}_k \right).$$

By construction, the leading-order solution contains only center modes; stable transients have been removed from the expansion. This choice is crucial for the RG procedure: all resonances are generated by combinations of the purely imaginary center exponents, and secular terms can occur only along center eigendirections.

### A.4 Definition of Secular and Bounded Terms

Following Chiba's RG formulation, we recursively define secular terms $R_i(C)$ and bounded corrections $h_t^{(i)}(C)$ at each order of the expansion. The secular terms govern the slow evolution of the center amplitudes, while the bounded corrections describe the embedding of the reduced dynamics into the full phase space. At first order, the secular term $R_1$ is defined as

$$R_1(C) = \lim_{t\to\infty}\frac{1}{t}\int_0^t V_c^* X^{-1}(s) g_1\left(X(s)U_c C\right) ds \tag{A.10}$$

and the corresponding bounded correction is

$$h_t^{(1)}(C) = X(t)\int^t \left[X^{-1}(s) g_1\left(X(s)U_c C\right) - U_c R_1(C)\right] ds \tag{A.11}$$

The unspecified lower limit of the integral indicates that we select particular solutions so as to preserve the elementary-term structure described below; any fixed lower limit leads to an equivalent RG system up to higher-order coordinate changes.

For higher orders, suppose that $R_k$ and $h_t^{(k)}$ have been defined for $1 \le k \le i-1$. The forcing term is first assembled $G_i$ from (A.3) using the lower-order solutions

$$x_0(t) = X(t)U_c C, \quad x_k(t) = h_t^{(k)}(C), \quad k \ge 1,$$

and then define

$$R_i(C) = \lim_{t\to\infty}\frac{1}{t}\int_0^t V_c^* X^{-1}(s)\left[G_i\left(X(s)U_c C, h_s^{(1)}(C), \dots, h_s^{(i-1)}(C)\right) - \sum_{k=1}^{i-1} Dh_s^{(k)}(C) R_{i-k}(C)\right]$$

(A.12)

$$h_t^{(i)}(C) = \int^t X(t)X^{-1}(s)\left[G_i(\cdots) - \sum_{k=1}^{i-1} Dh_s^{(k)}(C) R_{i-k}(C) - X(s)U_c R_i(C)\right] ds \tag{A.13}$$

The terms $Dh_s^{(k)}(C)R_{i-k}(C)$ arise from enforcing envelope invariance: the approximate solution must be independent of the arbitrary reference time at which the perturbation series is initialized. Differentiating the composite approximate solution with respect to this reference time generates contributions not only from explicitly secular terms (containing factors of $t$) but also from bounded corrections through their dependence on $C$. The subtractions in (A.12)-(A.13) are exactly those required to remove these spurious dependences and to obtain an autonomous reduced system for $C(t)$. Truncating at order $m$, the RG reduced equation and the associated RG mapping are

$$\begin{aligned} \dot{C} &= \sum_{i=1}^{m} \varepsilon^i R_i(C) \\ \Phi_t(C) &= X(t)U_c C + \sum_{i=1}^{m} \varepsilon^i h_t^{(i)}(C) \end{aligned} \tag{A.14}$$

The map $\Phi_t$ provides a local embedding of the reduced dynamics into the full phase space. For the theoretical results in Appendix B we will primarily use $\Phi_0$, the embedding at $t = 0$.

### A.5 Elementary-Term Structure and Computation

Under Assumption 3, the nonlinearity admits a representation in terms of symmetric multilinear tensors. Specifically, there exist symmetric $\ell$ -linear maps

$$L^{(\ell)} : \underbrace{\mathbb{R}^n \times \cdots \times \mathbb{R}^n}_{\ell \text{ times}} \to \mathbb{R}^n, \quad \ell \geq 2 \tag{A.15}$$

such that

$$g(x) = \sum_{\ell \geq 2} L^{(\ell)}(x, \ldots, x) \tag{A.16}$$

The derivative $D\,g(x)$ and higher derivatives $D^p g(x)$ are then sums of multilinear maps obtained from $\{L^{(\ell)}\}$ by inserting perturbations in the various argument slots. Consequently each forcing term $G_i$ in (A.3) is a finite linear combination of expressions of the form

$$L^{(\ell)}\left(x_{j_1},\ldots,x_{j_\ell}\right), \quad j_k \le i-1, j_1+\cdots+j_\ell = i-1. \tag{A.17}$$

By induction on $i$, using the explicit form of $x_0$ and the integral representation (A.15), one shows that each $x_i(t)$ can be expressed as a finite sum of elementary terms

$$p(C,\bar{C})e^{\lambda t}U_j \tag{A.18}$$

where:

- $U_j$ is either a center eigenvector or a stable eigenvector of A ;
- $p(C,\bar{C})$ is a polynomial in the complex amplitudes and their conjugates;
- $\lambda$ belongs to the integer module generated by the center eigenvalues, i.e.

$$\lambda = \sum_{k=1}^{N_c}\left(m_k - n_k\right)\mathrm{i}\omega_k, \quad m_k, n_k \in \mathbb{Z}_{\ge 0} \tag{A.19}$$

Moreover, each forcing term $G_i$ can be written as a sum of elementary terms of the form (A.23). This representation is preserved under partial differentiation with respect to $C$ and under the action of the tensors $L^{(\ell)}$.

Consider a single elementary term

$$G(t;C,\bar{C}) = p(C,\bar{C})e^{\lambda t}U_j$$

entering the right-hand side of the equation

$$\dot{x} = Ax + G(t;C,\bar{C}).$$

Let $\lambda_j$ denote the eigenvalue corresponding to $U_j$, so that

$$X(t)U_j = e^{\lambda_j t}U_j.$$

The Duhamel integral for a particular solution of the equation is

$$x(t) = \int_0^t X(t-s)G(s;C,\bar{C})ds \tag{A.20}$$

There are two distinct cases:

Case 1 (Resonant term). $\lambda = \lambda_j$.

By (A.20), we obtain

$$x(t) = tp(C,\bar{C})e^{\lambda_j t}U_j. \tag{A.21}$$

Thus the contribution of a resonant elementary term to the solution is secular (grows linearly in $t$), and the coefficient $p(C,\bar{C})$ contributes to the corresponding component of $R_i(C,\bar{C})$ via definition (A.17).

Case 2 (Non-resonant term). $\lambda \neq \lambda_j$.

Then (A.28) yields

$$x(t) = \frac{p(C,\bar{C})}{\lambda - \lambda_j}e^{\lambda t}U_j - \frac{p(C,\bar{C})}{\lambda - \lambda_j}e^{\lambda_j t}U_j. \tag{A.22}$$

Both contributions are bounded in time. The second term in (A.22) depends on our choice of lower integration limit and can be absorbed into the homogeneous solution. In the RG construction we keep only the first term in (A.22) as a particular solution; this choice preserves the elementary-term structure.

Remark. Case 1 is the unique mechanism that generates contributions to the secular terms $R_i(C,\bar{C})$. Since the leading-order solution $x_0$ contains only exponentials with purely imaginary exponents, the exponents $\lambda$ in (A.18) also have vanishing real part. However, for a stable $U_j$ eigenvector, the associated eigenvalue $\lambda_j$ has strictly negative real part, so the equality $\lambda = \lambda_j$ can never occur. Consequently, no secular terms arise in the stable eigendirections; all secular contributions lie in the center subspace and are captured by the projection $V_c^*$ in (A.10) and (A.12). In contrast, the bounded corrections $h_t^{(i)}$ may contain stable-mode components, which are essential for reconstructing the full-state dynamics.

Because the elementary-term structure is closed under differentiation and under the action of the multilinear tensors, the computation of higher-order RG equations reduces to applying the same recursive procedure to a finite list of elementary terms at each order.

## Appendix B. Theoretical Justification of RG Slow-Manifold Approximation and Hopf Persistence

### B.1 Setting and Main Statement

Consider a smooth autonomous system on $\mathbb{R}^n$ depending on a real parameter $\mu$, with an equilibrium at the origin for $\mu = \mu_0$. Denote the Jacobian at this equilibrium by $A$. We assume that $A$ satisfies the spectral hypotheses stated in the main text and Appendix A: there is a finite-dimensional center subspace $E^c$ associated with purely imaginary eigenvalues, and a complementary stable subspace $E^s$ associated with eigenvalues with strictly negative real parts. We also assume that there are no nontrivial Jordan blocks in either subspace.

We fix bases of right and left eigenvectors spanning $E^c$ and $E^s$, giving a block spectral decomposition of $A$ and a corresponding block expression for the linear flow. These are collected in formulas (B.1)-(B.2) below.

$$A = U \begin{bmatrix} \Lambda_c & 0 \\ 0 & \Lambda_s \end{bmatrix} V^*, \qquad U = [U_c \, U_s], \quad V^* = \begin{bmatrix} V_c^* \\ V_s^* \end{bmatrix} \tag{B.1}$$

$$V_c^* U_c = I, \qquad V_c^* U_s = 0, \qquad V_s^* U_c = 0, \qquad V_s^* U_s = I \tag{B.2}$$

The renormalization group (RG) construction, performed up to order $m$ in a small bookkeeping parameter $\varepsilon$, yields a reduced vector field on complex center amplitudes and a family of reconstruction maps. More precisely, one obtains:

1. A reduced system on $\mathbb{C}^{2Nc}$, written in terms of complex amplitudes $C$, which has the form displayed in (B.3). Its coefficients are constructed from the original nonlinearity via time-averaging formulas, as in (A.10) and (A.12).

$$\dot{C} = R(C,\mu) := \sum_{i=1}^{m} \varepsilon^i R_i(C,\mu), \tag{B.3}$$

2. A family of RG reconstruction maps $\Phi_t$ indexed by reference time $t$, which map center amplitudes back to the physical state space. The general form of $\Phi_t$ and the integral recursion defining its coefficients $h_t^{(i)}$ are given in (B.4)-(B.5).

$$\Phi_t(C,\mu) = X(t)U_c C + \sum_{i=1}^{m} \varepsilon^i h_t^{(i)}(C,\mu) \tag{B.4}$$

$$h_t^{(i)}(C,\mu) = X(t)\int_{t_0}^{t} X(s)^{-}\left[G_i\left(X(s)U_c C, h_s^{(1)}(C),\dots,h_s^{(i-1)}(C)\right) - \sum_{k=1}^{i-1} Dh_s^{(k)}(C)R_{i-k}(C)\right] - U_c R_i(C)ds$$

(B.5)

The real dynamics are recovered by imposing the complex-conjugacy constraints among amplitudes corresponding to positive and negative imaginary eigenvalues. These constraints define a real submanifold $\mathcal{M} \subset \mathbb{C}^{2N_c}$, described in (B.6). Restricting the RG reduced system and reconstruction to $\mathcal{M}$ gives a $2N_c$-dimensional real reduced system and a real embedding into $\mathbb{R}^n$

$$\mathcal{M} := \{ C \in \mathbb{C}^{2N_c} : C_{k+N_c} = \bar{C}_k, k = 1, \ldots, N_c \}. \tag{B.6}$$

The main result of this appendix has four parts:

1. The time-zero RG embedding $\Phi_0$ (as a special case of $\Phi_t$) defines an approximate invariant manifold tangent to $E^c$, and its image is real when restricted to $\mathcal{M}$.

2. The reduced dynamics induced on this manifold approximate the restriction of the full vector field up to order $\varepsilon^{m+1}$ in a suitable $C^r$ topology.

3. This approximate manifold is $C^r$-close, in the sense of Fenichel's theory, to a true normally hyperbolic invariant manifold for the full system.

4. If the RG-reduced system undergoes a nondegenerate Hopf bifurcation, then the full system also undergoes a Hopf bifurcation of the same qualitative type (supercritical or subcritical), with $O(\varepsilon^{m+1})$ accuracy in both the critical parameter and the first Lyapunov coefficient.

The rest of the appendix is organized as follows. Section B.2 proves invariance and reality of the RG manifold. Section B.3 shows that the RG manifold does not depend on the arbitrary choice of reference time. Section B.4 abstracts the situation to a general approximate invariant manifold, establishes the existence of a nearby true invariant manifold, and proves Hopf persistence. Section B. 5 specializes these abstract results back to the RG construction.

**B.2. Lemma 1: Invariance and Reality of the RG Manifold**

The complex RG reduced equations are defined on $\mathbb{C}^{2N_c}$. To obtain real dynamics, we impose a conjugacy constraint between amplitudes corresponding to eigenvalues $+i\omega_k$ and $-i\omega_k$. This defines the real constraint manifold $\mathcal{M}$ given in (B.6), which is a smooth real manifold of dimension

$2N_c$.

The RG reconstruction at a general reference time $t$ has a linear part given by the linear flow acting on center modes, plus a finite sum of nonlinear corrections $h_t^{(i)}$. This is recorded in (B.4). Each correction $h_t^{(i)}$ is bounded in time and satisfies an integral recursion in terms of lower-order corrections and the original nonlinearity, see (B.5). Evaluating this reconstruction at $t = 0$ defines the time-zero embedding $\mathbf{\Phi}_0$, whose explicit form is stated in (B.7).

$$\Phi_0(C,\mu) := U_c C + \sum_{i=1}^{m} \varepsilon^i h_0^{(i)}(C,\mu) \tag{B.7}$$

Lemma B. 1 states that the manifold $\mathcal{M}$ is invariant under the RG reduced dynamics and that the embedding $\mathbf{\Phi}_0$ is real valued when restricted to $\mathcal{M}$. Consequently, the image

$$M_{\mathrm{RG}} := \Phi_0(\mathcal{M}). \tag{B.8}$$

defined in (B.8), is a real $2\,\mathrm{N}_c$-dimensional submanifold of $\mathbb{R}^n$, tangent to the center subspace $E^c$ at the origin.

The proof is carried out by induction in the truncation order $m$. At leading order, the reduced equations are trivial and the reconstruction is purely linear, so the reality of $\mathbf{\Phi}_0$ on $\mathcal{M}$ follows directly from the conjugate-pair structure of the eigenvectors. Assuming that, up to order $m$, the reduced vector field preserves $\mathcal{M}$ and the corrections $h_0^{(i)}$ are real on $\mathcal{M}$, one inspects the recursion formula (B.5). The nonlinearity and its derivatives enter only through multilinear combinations that preserve conjugation when evaluated on $\mathcal{M}$. Using Wirtinger calculus for real-valued functions of complex variables, one verifies that this structure is preserved at order $m+1$. Hence the invariance of $\mathcal{M}$ and the reality of $\mathbf{\Phi}_0$ on $\mathcal{M}$ hold at all truncation orders.

### B.3 Lemma 2: Time Independence of the RG Manifold

For each reference time $t$, the RG reconstruction $\Phi_t$ induces a manifold

$$M_t := \Phi_t(\mathcal{M}) \subset \mathbb{R}^n$$

This manifold is defined by combining the linear flow with the bounded correction terms $h_t^{(i)}$, as in (B.4). Geometrically, the underlying RG manifold should not depend on the arbitrarily chosen reference time $t$, but only on the invariant center directions.

Using the block spectral decomposition of the linear flow summarized in (B.1)-(B.2), one shows that the linear part of $\Phi_t$ can be written as the linear flow acting on the initial center embedding. The corrections $h_t^{(i)}$ satisfy a covariance relation under time shifts: each $h_t^{(i)}$ is obtained from its time-zero counterpart $h_0^{(i)}$ by replacing the amplitudes $C$ with their linearly propagated values. This time-covariance property is encoded in (B.9).

$$h_t^{(i)}(C,\mu) = h_0^{(i)}\left(e^{\Lambda_c t}C,\mu\right) \tag{B.9}$$

Summing the linear part and the corrections, one obtains the relation (B.10),

$$\Phi_t(C,\mu) = X(t)U_c C + \sum_{i=1}^{m} \varepsilon^i h_t^{(i)}(C,\mu) = \Phi_0\left(e^{\Lambda_c t}C,\mu\right) \tag{B.10}$$

which explicitly expresses $\Phi_t$ in terms of $X(t)$, $\Phi_0$, and the linear center evolution. Since the linear center map preserves the constraint manifold $\mathcal{M}$, it follows that

$$M_t := \Phi_t(\mathcal{M}) = \Phi_0(X(t)\mathcal{M}) = \Phi_0(\mathcal{M}) = M_{\mathrm{RG}}. \tag{B.11}$$

as stated in (B.11). Thus the RG manifold is independent of the reference time $t$, and we may work with $M_{\mathrm{RG}}$ without loss of generality.

### B.4 Lemma 3: Approximate Manifold, Fenichel Persistence and Hopf Bifurcation

We now present an abstract version of the argument, which applies to any approximate invariant manifold satisfying suitable estimates. The RG manifold will later be shown to fit into this framework.

Consider a smooth family of vector fields on $\mathbb{R}^n$ depending on parameters $(\varepsilon, \mu)$,

$$\dot{z} = F(z; \varepsilon, \mu),$$

where $\varepsilon \geq 0$ is a small perturbation parameter and $\mu \in \mathbb{R}$ is a bifurcation parameter. Assume that for $\varepsilon = 0$ the linearization at the origin has a spectral decomposition

$$T_0\mathbb{R}^n = E^c \oplus E^s,$$

where $E^c$ is two-dimensional and corresponds to a simple pair of purely imaginary eigenvalues at $\mu = \mu_0$, while all eigenvalues on $E^s$ have strictly negative real parts bounded away from zero.

Suppose that for $\varepsilon > 0$ sufficiently small there exist:

1. A $C^r$ mapping

$$\Psi : B_\delta^{E^c} \to \mathbb{R}^n, \quad M_{\text{app}} := \left\{ z = \Psi(\theta) : \theta \in B_\delta^{E^c} \right\},$$

with $D\Psi(0) = U_c$ spanning the center subspace at $\varepsilon = 0$, so that $M_{\text{app}}$ is an approximate center(-stable) manifold.

2. A $C^r$ approximate reduced vector field

$$R_{\text{app}}(\theta, \mu), \quad \theta \in B_\delta^{E^c},$$

such that the invariance residual

$$\Delta(\theta,\mu;\varepsilon) := F(\Psi(\theta);\varepsilon,\mu) - D\Psi(\theta)R_{\text{app}}(\theta,\mu)$$

satisfies the uniform bound

$$\| \Delta(\cdot,\mu;\varepsilon)\|_{C^r\left(B_\delta^{E^c}\right)} \le C\varepsilon^{m+1}$$

for all $|\mu-\mu_0| \le \mu_*$ and all sufficiently small $\varepsilon$ , for some integer $m \ge 1$.

Assume further that for $\varepsilon = 0$ the approximate reduced system

$$\dot{\theta} = R_{\text{app}}(\theta,\mu)$$

undergoes a nondegenerate Hopf bifurcation at $(\theta,\mu) = (0,\mu_0)$ : its linearization has a simple pair of purely imaginary eigenvalues, the usual non-resonance and transversality conditions hold, and the first Lyapunov coefficient $l_1 \neq 0$.

Then there exists $\varepsilon > 0$ such that, for every $\varepsilon \in (0,\varepsilon_0]$ and $|\mu-\mu_0| \le \mu_*$ :

(i) The full system admits a locally invariant, normally hyperbolic $C^r$ manifold

$$M_\varepsilon = \left\{ z = \Psi(\theta) + U_s h_\varepsilon(\theta) : \theta \in B_{\delta_1}^{E^c} \right\},$$

with $\delta_1 > 0$ independent of $\varepsilon$ , and

$$\|h_\varepsilon\|_{C^r\left(B_{\delta_1}^{E^c}\right)} = O\left(\varepsilon^{m+1}\right)$$

Consequently,

$$\text{dist}_{C^r}\left(M_\varepsilon, M_{\text{app}}\right) = O\left(\varepsilon^{m+1}\right).$$

(ii) The reduced vector field on $M_\varepsilon$,

obtained by restricting the full dynamics to $M_\varepsilon$, satisfies

$$\left\| F_{\text{red}}(\cdot,\mu;\varepsilon) - R_{\text{app}}(\cdot,\mu) \right\|_{C^{r-1}\left(B_{\delta_1}^{E^c}\right)} = O\left(\varepsilon^{m+1}\right)$$

(iii) For $\varepsilon > 0$ sufficiently small, the full system exhibits a nondegenerate Hopf bifurcation on $M_\varepsilon$ at a parameter value $\mu = \mu_\varepsilon$ with

$$\mu_\varepsilon - \mu_0 = O\left(\varepsilon^{m+1}\right)$$

and the first Lyapunov coefficient $l_{1,\text{ full}}(\varepsilon)$ of the full system (restricted to $M_\varepsilon$ ) satisfies

$$l_{1,\text{full}}(\varepsilon) - l_1 = O\left(\varepsilon^{m+1}\right)$$

In particular, for all sufficiently small $\varepsilon$ , the sign of $l_{1,\text{ full}}(\varepsilon)$ coincides with the sign of $l_1$ , so the Hopf bifurcation of the full system has the same qualitative type (supercritical or subcritical) as that of the approximate reduced system.

Moreover, by symmetry of the construction, the converse holds: if the full system undergoes a nondegenerate Hopf bifurcation on a normally hyperbolic invariant manifold $M_\varepsilon$ of the form above, then for $\varepsilon$ sufficiently small the approximate reduced system on $M_{\text{app}}$ also exhibits a nondegenerate Hopf bifurcation with the same type, and the corresponding Hopf parameter values differ by $O\left(\varepsilon^{m+1}\right)$.

In what follows we suppress the explicit $\mu$ -dependence in most formulas for notational simplicity. One may regard $\mu$ as entering the linear part $A(\mu)$ so that the approximate reduced system at order $\varepsilon^0$ takes the form

$$R_{\text{app}}(\theta,\mu) = A_c(\mu)\theta + N_c(\theta,\mu),$$

and undergoes a Hopf bifurcation at $\mu = \mu_0$ . Fenichel's theorem is then applied to the corresponding suspended system in $(z,\mu)$, which poses no additional difficulty.

### B.4.1. Approximate Manifold and Invariance Residual

Let $M_{\text{app}}$ be a smooth manifold represented locally as a graph over $\mathbb{R}^{2N_c}$ by a mapping $\Psi$, as written in (B.12).

$$M_{\text{app}} := \{z = \Psi(\theta) : \theta \in \mathbb{R}^{2N_c}\} \tag{B.12}$$

We assume that $\Psi$ is tangent at the origin to the center subspace of the linearization, so that $M_{\text{app}}$ is an approximate center manifold.

Let $R_{\text{app}}(\theta, \mu)$ be a reduced vector field defined on the $\theta$ -coordinates, intended to approximate the restriction of the full dynamics to $M_{\text{app}}$ . The deviation from exact invariance is measured by the residual $\Delta(\theta)$ defined in (B.13).

$$\Delta(\theta) := F(\Psi(\theta), \mu_0) - D\Psi(\theta) R_{\text{app}}(\theta, \mu_0) \tag{B.13}$$

For the RG manifold, this residual is of order $\varepsilon^{m+1}$. In the abstract formulation, we assume that $\Delta$ satisfies a uniform bound as in (B.14)

$$\| \Delta(\theta) \|_{C^r} \le C \varepsilon^{m+1}. \tag{B.14}$$

on a fixed neighborhood and in a $C^r$ norm.

### B.4.2. Center-Stable Coordinates and Structure of the Pulled-Back System

Fix bases of the center and stable subspaces of the linearized system, and let $U_s$ be a matrix whose columns span the stable subspace. We introduce local coordinates $(\theta, \eta)$ by adding a stable component to the approximate manifold according to (B.15).

$$\Gamma(\theta, \eta) := \Psi(\theta) + U_s \eta, \qquad (\theta, \eta) \in \mathbb{R}^{2N_c} \times \mathbb{R}^{N_s} \tag{B.15}$$

For $(\theta,\eta)$ sufficiently close to the origin, this map is a diffeomorphism onto its image. Its Jacobian and inverse are denoted by block matrices $P_c$ and $P_s$, see (B.16)-(B.17).

$$D\Gamma(\theta,\eta)=\left[D\Psi(\theta),U_s\right]. \tag{B.16}$$

$$D\Gamma^{-1}(z)=\begin{bmatrix}P_c(\theta,\eta)\\ P_s(\theta,\eta)\end{bmatrix} \tag{B.17}$$

Tangency of $M_{\text{app}}$ to the center subspace implies the identities in (B.18).

$$D\Psi(0)=U_c \tag{B.18}$$

Pulling back the full vector field into $(\theta,\eta)$-coordinates yields the system written in (B.19).

$$\begin{bmatrix}\dot{\theta}\\ \dot{\eta}\end{bmatrix}=\begin{bmatrix}P_c(\theta,\eta)\\ P_s(\theta,\eta)\end{bmatrix}F(\Psi(\theta)+U_s\eta,\mu_0) \tag{B.19}$$

We expand the original vector field with respect to the stable coordinate $\eta$ near $\eta=0$. This yields a linear term in $\eta$ plus a remainder $N(\theta,\eta)$ that is at least quadratic in $\eta$, bounded as in (B.20).

$$F(\Psi(\theta)+U_s\eta,\mu_0)=F(\Psi(\theta),\mu_0)+DF(\Psi(\theta),\mu_0)U_s\eta+N(\theta,\eta),$$

$$\| N(\theta,\eta)\| \le C_N\| \eta\|^2 \tag{B.20}$$

Using the definition of the residual $\Delta(\theta)$ in (B.13), and decomposing the linear part into center and stable blocks denoted by $A_c$ and $\Lambda_s$, we can rewrite the system in the form (B.21)-(B.22).

$$\dot{\theta}=A_c\theta+Nc(\theta)+P_c(\theta,\eta)\Delta(\theta)+G(\theta,\eta) \tag{B.21}$$

$$\dot{\eta}=\Lambda_s\eta+P_s(\theta,\eta)\Delta(\theta)+B(\theta,\eta) \tag{B.22}$$

Here $A_c$ denotes the restriction of the linearization of the full vector field to the center subspace, expressed in the $\theta$ coordinates induced by $\Psi$. In particular, when the approximate manifold $M_{\text{app}}$ carries the approximate reduced dynamics $\dot{\theta} = R_{\text{app}}(\theta, \mu_0)$, we can write near $\theta = 0$

$$R_{\text{app}}(\theta, \mu_0) = A_c\theta + N_c(\theta),$$

where $N_c(\theta) = O(|\theta|^2)$. The term $\Delta(\theta)$ may contain linear contributions in $\theta$ accounting for the difference between the true linear part and the linearization of $R_{\text{app}}$; these are precisely incorporated via the projections $P_c\Delta$ and $P_s\Delta$. With this convention, the comparison between the global true and ideal systems below matches the comparison between the global vector field and its "ideal" reduced dynamics $A_c\theta + N_c(\theta)$ on the approximate center manifold.

The vector fields $G(\theta,\eta)$ and $B(\theta,\eta)$ collect all genuinely nonlinear contributions in $\eta$ and mixed $\theta$–$\eta$ interactions, after extracting the linear stable part $\Lambda_s\eta$. Crucially, we arrange that all linear dependence on $\eta$ sits in $\Lambda_s\eta$; in particular, we impose

$$G(\theta,0) = 0, \quad B(\theta,0) = 0, \quad D_\eta G(0,0) = 0, \quad D_\eta B(0,0) = 0, \tag{B.23}$$

for $\theta$ in a small neighborhood, as stated in (B.23).

Moreover, by shrinking the neighborhood if necessary, we can ensure the following growth estimates for some constant $L > 0$ :

$$\| G(\theta,\eta) \| \leq L(\| \theta \| + \| \eta \|) \| \eta \| \tag{B.24}$$

$$\| B(\theta,\eta) \| \leq L(\| \theta \| + \| \eta \|) \| \eta \| \tag{B.25}$$

as recorded in (B.24)-(B.25). These estimates encode two facts:

1. The nonlinear terms vanish when $\eta = 0$, so the approximate manifold has no spurious forcing in the normal direction in the ideal case.

2. The nonlinear terms are at least linear in $\eta$, reflecting that deviations from the manifold decay according to the stable linear dynamics, with higher-order corrections.

The contribution of the invariance residual $\Delta(\theta)$ to the right-hand side is controlled via the projection operators $P_c$ and $P_s$. Using smoothness of these operators and the bound (B.14), we obtain the estimate in (B.26).

$$\| P_c(\theta,\eta)\Delta(\theta)\| + \| P_s(\theta,\eta)\Delta(\theta)\| \le C'\varepsilon^{m+1}. \tag{B.26}$$

**B.4.3. Global Cutoff and Construction of Ideal and True Systems**

To apply Fenichel theory[22], it is convenient to extend the local system in $(\theta,\eta)$-coordinates to a globally defined system that coincides with the original one in a prescribed neighborhood and is globally Lipschitz. We do this by introducing a smooth cutoff function $\chi_\delta(\eta)$ depending only on the stable coordinate $\eta$, which equals one near $\eta=0$ and vanishes outside a ball of radius $2\delta$, with derivative bounded by a constant multiple of $1/\delta$. This is summarized in (B.27), and we define $\chi_\delta(\theta,\eta) = \chi_\delta(\theta)\chi_\delta(\eta)$.

$$\chi_\delta(\eta) = 1 \quad (\|\eta\| \le \delta), \qquad \chi_\delta(\eta) = 0 \quad (\|\eta\| \ge 2\delta), \qquad \|\nabla\chi_\delta(\eta)\| \le C_\chi / \delta, \tag{B.27}$$

Using $\chi_\delta$, we define two globally defined vector fields on the $(\theta,\eta)$-space:

1. The "global true system" obtained by including the residual terms $P_c\Delta$ and $P_s\Delta$ and multiplying all nonlinear and residual terms by $\chi_\delta(\eta)$. Its expression is given in (B.28).

$$\tilde{F}_{\text{global}} : \begin{cases} \dot{\theta} = A_c\theta + \chi_\delta(\theta,\eta)\left[P_c\Delta + G(\theta,\eta) + N_c(\theta)\right] \\ \dot{\eta} = \Lambda_s\eta + \chi_\delta(\theta,\eta)\left[P_s\Delta + B(\theta,\eta)\right] \end{cases} \tag{B.28}$$

2. The "global ideal system" obtained from the previous one by formally setting the residual to

zero, that is, by removing the $P_c\Delta$ and $P_s\Delta$ terms. This vector field is defined in (B.29).

$$\tilde{F}_{\text{global,ideal}} : \begin{cases} \dot{\theta} = A_c\theta + \chi_\delta(\theta,\eta)\left[N_c(\theta) + G(\theta,\eta)\right] \\ \dot{\eta} = \Lambda_s\eta + \chi_\delta(\theta,\eta)B(\theta,\eta) \end{cases} \tag{B.29}$$

By construction, both global systems coincide with the original pulled-back system (B.21)-(B.22) in the region where the cutoff equals one and the residual is small. Their difference is precisely the cutoff multiplied by the projected residual, as in (B.30),

$$F_{\text{global}}(\theta,\eta) - F_{\text{ideal}}(\theta,\eta) = \chi_\delta(\theta,\eta)\begin{bmatrix} P_c(\theta,\eta)\Delta(\theta) \\ P_s(\theta,\eta)\Delta(\theta) \end{bmatrix} \tag{B.30}$$

and hence is $C^r$ -small of order $\varepsilon^{m+1}$, according to (B.14) and smoothness of $P_c$, $P_s$. This is captured in (B.31).

$$\| F_{\text{global}} - F_{\text{ideal}} \|_{C^r} \leq K\varepsilon^{m+1}. \tag{B.31}$$

**B.4.4. Verification of Fenichel's Hypotheses for the Ideal System**

We now check that the global ideal system satisfies the conditions of Fenichel's persistence theorem.

On the hyperplane $\eta=0$, the $\eta$ -equation of the ideal system reduces to

$$\dot{\eta} = \Lambda_s\eta + \chi_\delta(\theta,0)B(\theta,0).$$

Because we arranged that $B(\theta,0)=0$ for all sufficiently small $\theta$, this reduces to $\dot{\eta} = \Lambda_s\eta$ on $\eta=0$. Thus the manifold $\eta=0$ is an exact invariant manifold for the ideal system.

The Jacobian of the ideal system in the normal direction at points on $\eta=0$ is given by $D_\eta F_{\text{ideal}}(\theta,0)$. Using the spectral gap of $\Lambda_s$ and the assumption $D_\eta B(0,0)=0$, we obtain the uniform spectral bound in (B.32):

$$\Re\,\sigma\left(D_\eta F_{\text{ideal}}(\theta,0)\right) \le -\beta < 0 \quad \text{for } \|\theta\| \le \delta \tag{B.32}$$

the real parts of all eigenvalues of the normal Jacobian lie to the left of some negative constant $-\beta$, independent of $\theta$ in a neighborhood of the origin. This implies normal hyperbolicity of the invariant manifold $\eta=0$.

The tangential dynamics on $\eta=0$ are governed by the reduced $\theta$ -equation of the ideal system, which reads

$$\dot{\theta} = A_c\theta + \chi_\delta(\theta,0)N_c(\theta) + \chi_\delta(\theta,0)G(\theta,0).$$

Since $G(\theta,0)=0$ in our setup, the tangential dynamics near the origin are dominated by the center block $A_c$, possibly plus higher-order corrections in $\theta$ .

Finally, the difference between the global true and ideal systems is $C^r$ -small and bounded by $K\varepsilon^{m+1}$ for some constant $K$, as stated in (B.31). Therefore, all hypotheses of Fenichel's theorem are satisfied for the pair consisting of the ideal system and its $C^r$ -small perturbation given by the global true system.

Remark: By shrinking the neighborhood in the $(\theta,\eta)$ -coordinates if necessary, we may and do fix a radius $\delta > 0$ such that the above spectral bound holds uniformly for all $\theta$ with $\|\theta\| \le \delta$. In particular, $\delta$ can be chosen independently of $\varepsilon$ . Although the RG reconstruction map $\Phi$ and hence the local coordinate change $\Gamma$ may contain $\varepsilon$ -dependent higher-order terms, these terms only modify the embedding on a ball whose radius is of order $\varepsilon$ and therefore remain well inside the fixed ball of radius $\delta$ for all sufficiently small $\varepsilon$ . Equivalently, one may assume that $\Phi$ consists of a linear part plus an $\varepsilon$ -small nonlinear correction, so that $\Gamma$ remains a diffeomorphism on the $\delta$ -ball uniformly in $\varepsilon$ . This uniform choice of $\delta$ ensures that the spectral gap hypothesis entering Fenichel's theorem is satisfied on a common neighborhood for all small $\varepsilon$ , the spectral gap can be made sufficiently large to achieve the desired strength of normal hyperbolicity; for a detailed discussion, see [23].

#### B.4.5. Existence and Approximation of a True Invariant Manifold

Applying Fenichel's persistence theorem, we conclude that there exists a unique invariant manifold for the global true system, written as a graph in a neighborhood of $\eta=0$. This manifold depends smoothly on $\varepsilon$ , and its graph satisfies a $C^r$ estimate of order $\varepsilon^{m+1}$ [23] , as recorded in (B.33).

$$\eta = h_\varepsilon(\theta), \qquad \| h_\varepsilon\|_{C^r} = \mathcal{O}(\varepsilon^{m+1}), \tag{B.33}$$

The reduced vector field on this invariant graph is obtained by substituting $\eta=h_\varepsilon(\theta)$ into the $\theta$ -equation of the global true system, giving $F_{\text{red}}\left(\theta,\mu_0\right)$ defined in (B.34).

$$F_{\text{red}}(\theta,\mu_0) := A_c\theta + N_c(\theta) + G\big(\theta, h_\varepsilon(\theta)\big) + P_c\big(\theta, h_\varepsilon(\theta)\big)\Delta(\theta) \tag{B.34}$$

Using the residual bound (B.14), the growth estimates (B.24),(B.26), and the(B.33) estimate for $h_\varepsilon$, one shows that the reduced vector field $F_{\text{red}}$ is $C^{r-1}$ -close to the approximate reduced field $R_{\text{app}}\left(\theta,\mu_0\right)$, with an error of order $\varepsilon^{m+1}$. This is expressed in (B.35).

$$\| F_{\text{red}} - R_{\text{app}}\|_{C^{r-1}} = \mathcal{O}(\varepsilon^{m+1}) \tag{B.35}$$

Inside the region where the cutoff function equals one, the global true system coincides with the original system in $(\theta,\eta)$ coordinates. Pulling back the invariant graph via the diffeomorphism $\Gamma$ defined in (B.15), we obtain a local invariant manifold of the original system, for $\theta$ in a small neighborhood, as written in (B.36).

$$M_{\varepsilon} = \left\{ \Psi(\theta) + U_s h_{\varepsilon}(\theta) : \| \theta \| \le \delta \right\} \tag{B.36}$$

This manifold is normally hyperbolic and $C^r$ -close to the approximate manifold $M_{\text{app}}$ . The distance estimate is given in (B.37).

$$\text{dist}_{C^r}(M_{\varepsilon}, M_{\text{app}}) = \mathcal{O}(\varepsilon^{m+1}). \tag{B.37}$$

Remark: We emphasize the role of the radius $\delta$ in this construction. First, $\delta$ is used to localize the coordinate chart $\Gamma$ and to ensure that all mappings involved, in particular the RG reconstruction map $\Phi$ in Appendix A, remain diffeomorphisms on a fixed neighborhood of the origin. Second, $\delta$ controls the support of the cutoff used to splice the local dynamics with a global Lipschitz field, and hence the region on which Fenichel's theorem is applied. As noted above, one can choose \delta>0 independently of $\varepsilon$ by viewing the $\varepsilon$ -dependent terms in $\Phi$ and $\Gamma$ as small nonlinear corrections to a fixed linear isomorphism. For all sufficiently small $\varepsilon$ , the spectral gap of the stable block $\Lambda_s$ then holds uniformly on the $\delta$ ball, so that the same $\delta$ can be used throughout the argument.

**B.4.6. Hopf Bifurcation Persistence and First Lyapunov Coefficient**

We now incorporate parameter dependence and Hopf bifurcation. Suppose that the approximate reduced system

$$\dot{\theta} = R_{\text{app}}(\theta, \mu)$$

undergoes a nondegenerate Hopf bifurcation at $\theta = 0$ and $\mu = \mu_0$ . This means that its linearization at the equilibrium has a simple pair of purely imaginary eigenvalues, the usual non-resonance and transversality conditions are satisfied, and the first Lyapunov coefficient $l_1$ is nonzero. Define a homotopy for the reduced vector field on the invariant manifold by interpolating between the approximate reduced field and the true reduced field:

$$F_\tau^{\text{red}}(\theta,\mu) = (1-\tau)R_{\text{app}}(\theta,\mu) + \tau F_{\text{red}}(\theta,\mu) \tag{B.38}$$

For $\tau \in [0,1]$, as stated in (B.38).For each $\tau$ , denote by $\lambda(\theta,\mu,\tau)$ the critical eigenvalue of the linearization of $F_\tau^{\text{red}}$ at its equilibrium (the eigenvalue whose real part is closest to zero).We collect the Hopf conditions in a mapping $\mathcal{H}$ defined in (B.39),

$$\mathcal{H}(\theta,\mu,\omega,\tau) = \left(F_\tau^{\text{red}}(\theta,\mu), \Re\lambda(\theta,\mu,\tau), \Im\lambda(\theta,\mu,\tau) - \omega\right) \tag{B.39}$$

which takes as arguments $\theta,\mu$ , a positive angular frequency $\omega$ , and the homotopy parameter $\tau$ , and returns the combination of equilibrium and spectral conditions characterizing a Hopf point: vanishing of the vector field, vanishing of the real part of the critical eigenvalue, and matching of its imaginary part with $\omega$ .

At $\tau = 0$ , the mapping $\mathcal{H}$ describes the Hopf bifurcation of $R_{\text{app}}$ . The nondegeneracy of this Hopf bifurcation ensures that the Jacobian of $\mathcal{H}$ with respect to $(\theta,\mu,\omega)$ is invertible at the reference point $\left(0,\mu_0,\omega_0,0\right)$ . The $C^{r-1}$ -closeness of $F_{\text{red}}$ and $R_{\text{app}}$ for the relevant values of $\theta$ and $\mu$ , as quantified in (B.35), implies that $F_\tau^{\text{red}}$ depends smoothly on $\tau$ and that the mapping $\mathcal{H}$ is a small $C^2$ perturbation of $\mathcal{H}(\cdot,\cdot,\cdot,0)$ . By the Implicit Function Theorem, there thus exists a smooth family $(\theta(\tau),\mu(\tau),\omega(\tau))$ solving $\mathcal{H} = 0$ for $\tau$ in a neighborhood of zero, which can be extended to $\tau = 1$ for sufficiently small $\varepsilon$ . In particular, at $\tau = 1$ we obtain a Hopf bifurcation point $(\theta(1),\mu(1),\omega(1))$ for the reduced dynamics $F_{\text{red}}$ on the true invariant manifold

$M_\varepsilon$. The parameter shift satisfies

$$\mu(1) = \mu_0 + O\left(\varepsilon^{m+1}\right) \tag{B.40}$$

as stated in (B.40).

The first Lyapunov coefficient for the full system restricted to $M_\varepsilon$, denoted $l_{1,\text{ full}}$, depends smoothly on the derivatives of the reduced vector field up to third order. Since $F_{\text{red}}$ and $R_{\text{app}}$ are $C^3$-close (provided $r \geq 4$), we obtain

$$l_{1,\text{full}} = l_1 + O\left(\varepsilon^{m+1}\right) \tag{B.41}$$

as in (B.41). For sufficiently small $\varepsilon$, the sign of $l_{1,\text{ full}}$ agrees with that of $l_1$, so the Hopf bifurcation of the full system has the same type (supercritical or subcritical) as that of the approximate reduced system and its classical normal form.

**B.5 Completion of the Main Theorem for the RG Construction**

We now apply the abstract framework of Section B. 4 to the RG manifold constructed in Appendix A.

First, by Lemma B. 1 and the explicit RG formulas, the constraint manifold $\mathcal{M}$ is invariant for the RG reduced system, and the time-zero reconstruction $\Phi_0$ is real on $\mathcal{M}$. Hence

$$M_{\text{app}} := M_{\text{RG}} = \Phi_0(\mathcal{M})$$

is a real embedded submanifold tangent to the center subspace of the full system at the origin.

Second, we construct the approximate manifold via the RG map and approximate vector field via the RG flow $F_M$ and confirm that $\left\|F_M(\theta) - F(\theta)\right\| \sim O(\epsilon^{m+1})$ ,Let $\varphi_t^{(m)}$ denote the flow generated by the RG vector field (A12). Then the following equation defines a vector field $F_{M,t}$ from $M_{app}$ to $\mathbb{R}^n$.

$$F_{M,t}: M_{app} \to \mathbb{R}^n, \theta \mapsto \left.\frac{d\Psi_{a,t}(\theta)}{da}\right|_{a=t} = \left.\frac{d}{da}\alpha_a\left(\varphi_{a-t}^{(m)}\left(\alpha_t^-(\theta)\right)\right)\right|_{a=t}$$

This vector field is in fact time-independent. Using the relations (B.42)

$$R_i\left(e^{\Lambda_C t}C\right) = e^{\Lambda_C t}R_i(C), \quad h_t(C) = h_{t'}\left(e^{\Lambda_C\left(t-t'\right)}C\right) \tag{B.42}$$

direct calculation as following yields

$$\begin{aligned}
&\left.\frac{d\Psi_{a,t+t_0}}{da}\right|_{a=t+t_0} = \left.\frac{d}{da}\alpha_a\left(\varphi_{a-t-t_0}^{(m)}\left(\alpha_{t+t_0}^-(\theta)\right)\right)\right|_{a=t_0} \\
&= \left(\frac{\partial}{\partial a}\alpha\right)_{t+t_0}\left(\alpha_{t+t_0}^-(\theta)\right) + D\alpha_{t_0+t}\left(\frac{d\varphi^{(m)}}{da}\right)\left(\alpha_{t+t_0}^-(\theta)\right) \\
&= \left(\frac{\partial}{\partial a}\alpha\right)_t\left(e^{\Lambda_t t_0}\alpha_{t+t_0}^-(\theta)\right) + D\alpha_t e^{\Lambda_t t_0}\left(\frac{d\varphi^{(m)}}{da}\right)\left(\alpha_{t+t_0}^-(\theta)\right) \\
&= \left(\frac{\partial}{\partial a}\alpha\right)_t\left(\alpha_t^-(\theta)\right) + D\alpha_t\left(\frac{d\varphi^{(m)}}{da}\right)\left(e^{\Lambda_c t_0}\alpha_{t+t_0}^-(\theta)\right) = \left(\frac{\partial}{\partial a}\alpha\right)_t\left(\alpha_t^-(\theta)\right) + D\alpha_t\left(\frac{d\varphi^{(m)}}{da}\right)\left(\alpha_t^-(\theta)\right) = \left.\frac{d\Psi_{a,t}}{da}\right|_{a=t}
\end{aligned}$$

establishing time-independence.

We now verify that $\left\|F_M(\theta) - F(\theta)\right\| \sim O(\epsilon^{m+1})$. Setting $t = t_0$, we compute $F_M(\theta)$ as following

$$
\begin{aligned}
&\frac{d}{da}\alpha_a\left(\varphi_{a-t}^{(m)}\left(\alpha_t^-(\theta)\right)\right)=\left(\frac{\partial}{\partial a}\alpha\right)_t\left(\alpha_t^-(\theta)\right)+D\alpha_{\alpha_t^-}\frac{d\varphi^{(m)}}{da}\left(\alpha_t^-(\theta)\right)\\
&=\left(\frac{\partial}{\partial a}\left(X(a)U_c\alpha_t^-(\theta)+h_a^{(1)}\left(\alpha_t^-(\theta)\right)+\cdots+{}^m h_a^{(m)}\left(\alpha_t^-(\theta)\right)\right)+D\alpha_{t_{\alpha_t^-(\theta)}}\left(\sum_{i=1}^{m}\varepsilon^i R_i\left(\alpha_t^-(\theta)\right)\right)\right.\\
&=AX(t)U_c\alpha_t^-(\theta)+\sum_{i=1}^{m}\epsilon^i\left(\begin{array}{c}Ah_t^{(i)}\left(\alpha_t^-(\theta)\right)+G_i\left(X(s)U_c\alpha_t^-(\theta),h_s^{(1)}\left(\alpha_t^-(\theta)\right),\cdots,h_s^{(i-1)}\left(\alpha_t^-(\theta)\right)\right)\\ -\sum_{k=1}^{i-1}\left(Dh_s^{(k)}\right)_{\alpha_t^-(\theta)}R_{i-k}\left(\alpha_t^-(\theta)\right)-X(t)U_cR_i\left(\alpha_t^-(\theta)\right)\end{array}\right)\\
&+\left(X(t)U_c+\sum_{i=1}^{m}\epsilon^i Dh_t^{(i)}\right)\left(\sum_{i=1}^{m}\varepsilon^i R_i\left(\alpha_t^-(\theta)\right)\right)\\
&=A\alpha_t\left(\alpha_t^-(\theta)\right)+\sum_{i=1}^{m}\epsilon^i G_i\left(X(s)U_c\alpha_t^-(\theta),h_s^{(1)}\left(\alpha_t^-(\theta)\right),\cdots,h_s^{(i-1)}\left(\alpha_t^-(\theta)\right)\right)+O\left(\varepsilon^{m+1}\right)
\end{aligned}
$$

The $O(\varepsilon^{m+1})$ terms, while difficult to write explicitly, can be shown to be bounded on compact sets by the assumptions and the structure of $\alpha$ and $R$. Meanwhile, $F\left(\Psi\left(\theta\right)\right)$ is computed as following

$$
\begin{aligned}
F(\theta)&=F\left(\alpha\left(\alpha^-(\theta)\right)\right)=A\theta+\varepsilon g_1\left(\alpha\left(\alpha^-(\theta)\right)\right)+\varepsilon^2 g_2\left(\alpha\left(\alpha^-(\theta)\right)\right)+\varepsilon^3 g_3\left(\alpha\left(\alpha^-(\theta)\right)\right)\\
&+\varepsilon^4 g_4\left(\alpha\left(\alpha^-(\theta)\right)\right)+\varepsilon^5 g_5\left(\alpha\left(\alpha^-(\theta)\right)\right)\\
&=A\theta+\sum_{i=1}^{m}\epsilon^i G_i\left(X(s)U_c\alpha_t^-(\theta),h_s^{(1)}\left(\alpha_t^-(\theta)\right),\cdots,h_s^{(i-1)}\left(\alpha_t^-(\theta)\right)\right)+O\left(\varepsilon^{m+1}\right)
\end{aligned}
.
$$

Comparing the expansions above yields the invariance residual estimate (B.14) for the RG manifold; consequently (B.35) holds Therefore, the pair $(M_{\mathrm{RG}},R)$ satisfies the hypotheses of Section B. 4 with $R_{app}=F_M$.

Third, applying the abstract Lemma 3 to this pair, we obtain a true normally hyperbolic invariant manifold $M_\varepsilon$ of the original system, which is $C^r$-close to $M_{\mathrm{RG}}$ with distance of order $\varepsilon^{m+1}$, as in (B.36)-(B.37). The reduced vector field $F_{\mathrm{red}}$ on $M_\varepsilon$ is $C^{r-1}$-close to the RG reduced vector field $R$, with the same order of accuracy (B.35). Normal hyperbolicity of $M_\varepsilon$ follows from the stable spectrum of $\Lambda_s$ .

Finally, if the RG reduced system exhibits a nondegenerate Hopf bifurcation at some parameter $\mu_0$, then by Section B.4.6 this Hopf bifurcation persists on the true invariant manifold $M_\varepsilon$ . The critical parameter for the Hopf point of the full system differs from that of the RG reduced system by $O\left(\varepsilon^{m+1}\right)$, and the corresponding first Lyapunov coefficient differs by the same order, as expressed in (B.40)-(B.41). Consequently, the full system undergoes a Hopf bifurcation of the same qualitative type as the RG reduced system.

This completes the proof of the main statement on the validity of the RG slow-manifold approximation and the persistence of Hopf bifurcations.

**Appendix C. Relation of the RG Hopf coefficients to the classical center-manifold normal form**

This appendix briefly clarifies how the coefficients appearing in the RG amplitude equation are related to the standard coefficients obtained from center-manifold reduction and Hopf normal-form theory. The purpose is not to reproduce the full normal-form derivation, but to show, in the notation of the present paper, that the RG coefficient used in the numerical sections is the same cubic quantity that determines the local Hopf criticality[5], up to the conventional scaling associated with the choice of complex amplitude.

For simplicity, consider a one-parameter family of systems near a simple Hopf point,

$$\dot{x} = A(\mu)x + \frac{1}{2}B(x,x) + \frac{1}{6}C(x,x,x) + \cdots, \tag{C.1}$$

Where $x \in \mathbb{R}^n$ , the bilinear and trilinear maps $B$ and $C$ are symmetric, and $\mu = \mu_0$ is a Hopf point of the linearized system. Let

$$A_0 := A(\mu_0), \qquad A_\mu := \frac{\partial A}{\partial \mu}(\mu_0).$$

We assume that $A_0$ has a simple critical pair $\pm i\omega$ with right eigenvector $V_c$ and left eigenvector $W_c$ normalized by

$$A_0 V_c = i\omega V_c, \qquad W_c^* A_0 = i\omega W_c^*, \qquad W_c^* V_c = 1.$$

Following the near-critical scaling used in the main text, we set

$$\mu = \mu_0 + \varepsilon^2 \delta\mu, \qquad x = \varepsilon y. \tag{C.2}$$

Then $C1$ becomes

$$\dot{y} = A_0 y + \varepsilon \frac{1}{2} B(y, y) + \varepsilon^2 \left( A_\mu \, \delta\mu \, y + \frac{1}{6} C(y, y, y) \right) + O(\varepsilon^3). \tag{C.3}$$

In the notation of Appendix A, this corresponds to

$$g_1(y) = \frac{1}{2} B(y, y), \qquad g_2(y) = A_\mu \, \delta\mu \, y + \frac{1}{6} C(y, y, y).$$

At leading order we take a center-only solution,

$$y_0 = V_c C e^{i\omega t} + \bar{V}_c \bar{C} e^{-i\omega t}, \tag{C.4}$$

where $C$ is the complex center amplitude. Since $g_1(y_0)$ contains only the harmonics $2i\omega$, $0$, and $-2i\omega$, there is no first-order resonance at $e^{\pm i\omega t}$, and therefore

$$R_1 = 0. \tag{C.5}$$

A convenient particular solution of the first-order problem is

$$y_1 = \frac{1}{2}(2i\omega I - A_0)^{-1} B(V_c, V_c) C^2 e^{2i\omega t} - A_0^{-1} B(V_c, \bar{V}_c) C\bar{C} + \frac{1}{2}(-2i\omega I - A_0)^{-1} B(\bar{V}_c, \bar{V}_c) \bar{C}^2 e^{-2i\omega t}.$$

(C.6)

The cubic contribution to the resonant term at order $\varepsilon^2$ comes from

$$Dg_1(y_0)y_1 + g_2(y_0).$$

Projecting the coefficient of $e^{i\omega t}$ onto the critical adjoint direction $W_c^*$ gives the RG amplitude equation

$$\dot{C} = \varepsilon^2\left(\sigma\,\delta\mu\, C + \gamma C\,|\,C\,|^2\right) + O(\varepsilon^3), \tag{C.7}$$

with

$$\sigma = W_c^* A_\mu V_c, \tag{C.8}$$

And

$$\gamma = W_c^*\left[\frac{1}{2}C(V_c,V_c,\bar{V}_c) - B\left(V_c, A_0^{-1}B(V_c,\bar{V}_c)\right) + \frac{1}{2}B\left(\bar{V}_c,(2i\omega I - A_0)^{-1}B(V_c,V_c)\right)\right]. \tag{C.9}$$

Equation $C7$ is precisely the local complex Hopf amplitude equation used throughout the paper. Its linear coefficient $\sigma$ gives the projected sensitivity of the critical eigenvalue with respect to the unfolding parameter, while $\Re(\gamma)$ determines the local Hopf criticality and $\Im(\gamma)$ gives the leading nonlinear frequency correction.

Formula $C9$ is the same cubic combination that appears in classical center-manifold/normal-form calculations for a simple Hopf bifurcation, up to notational conventions and amplitude scaling. In particular, the first Lyapunov coefficient $l_1$ is proportional to $\Re(\gamma)$; more precisely, under the common convention in which the reduced equation is written as

$$\dot{C} = i\omega C + \sigma\delta\mu C + \gamma C\,|\,C\,|^2 + \cdots,$$

one has

$l_1 = \frac{1}{\omega}\Re(\gamma)$ up to the sign/scaling convention adopted for the complex amplitude.

(C.10)

Thus, for the purposes of local bifurcation classification, the RG cubic coefficient carries exactly the same information as the classical Hopf normal-form coefficient.

This observation also explains the decomposition used in the numerical examples. If the physical nonlinear restoring term is additively decomposed, for example as

$$f^{(3)}(x) = k_h^3 f_h^{(3)}(x) + k_\alpha^3 f_\alpha^{(3)}(x) + k_\beta^3 f_\beta^{(3)}(x),$$

then the induced trilinear form $C(\cdot,\cdot,\cdot)$ is additive in the same way, and hence the reduced cubic coefficient $\gamma$ inherits the same additivity. By contrast, a quadratic nonlinearity may contribute to $\gamma$ through the slaving term $y_1$, i.e. through noncritical-mode mediation. This is exactly the mechanism discussed in Discussion III for the $k_{h\alpha}$ -type coupling: its effect on the Hopf cubic coefficient is not a direct physical cubic stiffness, but an induced cubic contribution generated by lower-order excitation and feedback of noncritical directions.

**Appendix D Definition of the augmented airfoil-section model**

This appendix summarizes the airfoil-section model used in the numerical examples. The purpose is to make the state variables, governing equations, and coefficient definitions explicit, while keeping the main text focused on the RG reduction itself. Standard derivations of the underlying unsteady aerodynamic model may be found in the aeroelasticity literature[24] and are therefore not repeated here.

**Structural coordinates and dimensional parameters**

We consider a classical binary airfoil section with three structural coordinates: plunge $h$ , pitch $\alpha$ ,

and control-surface deflection $\beta$. In addition, one aerodynamic memory state $z$ is introduced to represent the circulatory lag associated with unsteady aerodynamic loading in the time domain. The generalized coordinate vector is therefore

$$q=[h,\alpha,\beta,z]^{\mathrm{T}}.$$

The main geometric and inertial parameters are the half-chord $b$, the dimensionless elastic-axis location $a$, and the dimensionless hinge location $c$. The quantities $m_t$ and $m_w$ denote, respectively, the total mass of the airfoil-support system and the mass associated with the wing--control-surface subsystem. The static moments and inertia-related parameters are written in the standard form

$$x_a=\frac{S_\alpha}{m_w b},\qquad x_\beta=\frac{S_\beta}{m_w b},\qquad r_\alpha^2=\frac{I_\alpha}{m_w b^2},\qquad r_\beta^2=\frac{I_\beta}{m_w b^2},$$

where $S_\alpha$ is the static moment of the wing--control-surface system about the elastic axis, $S_\beta$ is the static moment of the control surface about the hinge, and $I_\alpha, I_\beta$ are the corresponding mass moments of inertia. Structural damping is introduced through a Rayleigh-type[25] construction in the structural part of the model. The explicit coefficients used in computation are included below through the augmented matrices.

**Augmented governing equations**

With the aerodynamic state included, the governing equations are written as

$$M\ddot{q}+C\dot{q}+Kq+F_{\mathrm{nl}}(q)=0, \tag{D.1}$$

where $M$, $C$, and $K$ are $4\times 4$ matrices. The nonlinear term is structural and, in the examples considered in the main text, has the form

$$F_{\mathrm{nl}}(q)=\begin{bmatrix} m_w\omega_h^2 k_{h3}h^3 \\ m_w b^2 r_\alpha^2\omega_\alpha^2 k_{\alpha 3}\alpha^3 \\ m_w b^2 r_\beta^2\omega_\beta^2 k_{\beta 3}\beta^3 \\ 0 \end{bmatrix}. \tag{D.2}$$

The corresponding first-order state used in the RG reduction is

$$X=\begin{bmatrix} q \\ \dot{q} \end{bmatrix}=\left[h,\alpha,\beta,z,\dot{h},\dot{\alpha},\dot{\beta},\dot{z}\right]^{\mathrm{T}},$$

which leads to

$$\dot{X}=\begin{bmatrix} 0 & I \\ -M^{-1}K & -M^{-1}C \end{bmatrix}X+\begin{bmatrix} 0 \\ -M^{-1}F_{\mathrm{nl}}(q) \end{bmatrix}. \tag{D.3}$$

This is the form used in the main text for linearization, Hopf detection, and RG coefficient extraction.

**Aerodynamic load representation and Wagner approximation**

The unsteady aerodynamic loads are modeled using an incompressible Theodorsen-type formulation[26] and rewritten in time domain through a Wagner-function approximation[27][28][29]. Let $\alpha_e$ denote the effective angle of attack. In the notation used here,

$$\alpha_e=\alpha+\frac{\dot{h}}{U}+\frac{b}{U}\left(\frac{1}{2}-a\right)\dot{\alpha}+\frac{T_{10}}{\pi}\beta+\frac{b}{2\pi U}T_{11}\dot{\beta}. \tag{D.4}$$

The unsteady effect contribution is represented through the Wagner kernel $\varphi(t)$. Using a two-exponential approximation, we write

$$\varphi(t)\approx c_0-c_1e^{-c_3t}-c_2e^{-c_4t}, \tag{D.5}$$

With[26]

$$c_0=1,\qquad c_1=0.165,\qquad c_2=0.0455,\qquad c_3=0.335,\qquad c_4=0.3. \tag{D.6}$$

Introducing the aerodynamic state $Z$ yields a finite-dimensional realization of the circulatory lag. In the present notation, the state equation is written compactly as

$$\ddot{z} = -c_3 c_4 \Big(\frac{U}{b}\Big)^2 z - (c_3 + c_4)\frac{U}{b}\dot{z} + \frac{U}{b}\alpha_e. \tag{D.7}$$

Equivalently, after substituting (D.4), one obtains the same augmented linear coupling used in the matrices below. The precise realization is not essential for the RG derivation; what matters is that the aerodynamic memory effect is represented by a linear augmented state and therefore enters the local reduction through the linear operator.

**Theodorsen-related constants**

For completeness, the auxiliary constants appearing in the aerodynamic matrices are listed here[26]:

$$\begin{aligned}
T_1 &= -\frac{2+c^2}{3}\sqrt{1-c^2} + c\arccos(c),\\
T_3 &= -\frac{1-c^2}{8}(5c^2+4) + \frac{c(7+2c^2)\sqrt{1-c^2}\arccos(c)}{4} - \Big(\frac{1}{8}+c^2\Big)\big(\arccos(c)\big)^2,\\
T_4 &= c\sqrt{1-c^2} - \arccos(c),\\
T_5 &= -(1-c^2) - \big(\arccos(c)\big)^2 + 2c\sqrt{1-c^2}\arccos(c),\\
T_7 &= \frac{c(7+2c^2)\sqrt{1-c^2}}{8} - \Big(\frac{1}{8}+c^2\Big)\arccos(c),\\
T_8 &= -\frac{1}{3}(1+2c^2)\sqrt{1-c^2} + c\arccos(c),\\
T_9 &= \frac{1}{2}\left(\frac{(1-c^2)^{3/2}}{3} + aT_4\right),\\
T_{10} &= \sqrt{1-c^2} + \arccos(c),\\
T_{11} &= (2-c)\sqrt{1-c^2} + (1-2c)\arccos(c),\\
T_{12} &= (2+c)\sqrt{1-c^2} - (1+2c)\arccos(c),\\
T_{13} &= \frac{1}{2}\big[-T_7 - (c-a)T_1\big].
\end{aligned}$$

**Augmented matrices used in computation**

The matrices in (D.3) are written as follows.

First, the augmented mass matrix is

$$M = \begin{bmatrix} m_t + \pi\rho b^2 & S_\alpha - a\pi\rho b^3 & S_\beta - \rho b^3 T_1 & 0 \\ S_\alpha - a\pi\rho b^3 & I_\alpha + \pi\left(\frac{1}{8} + a^2\right)\rho b^4 & I_\beta + S_\beta (c-a)b - \rho b^4\left(T_7 + (c-a)T_1\right) & 0 \\ S_\beta - \rho b^3 T_1 & I_\beta + S_\beta (c-a)b + 2\rho b^4 T_{13} & I_\beta - \frac{\rho b^4}{\pi} T_3 & 0 \\ 0 & 0 & 0 & 1 \end{bmatrix}.$$

The damping matrix is

$$C = \begin{bmatrix} C_{11} & C_{12} & C_{13} & C_{14} \\ C_{21} & C_{22} & C_{23} & C_{24} \\ C_{31} & C_{32} & C_{33} & C_{34} \\ C_{41} & C_{42} & C_{43} & C_{44} \end{bmatrix},$$

with

$$\begin{aligned}
C_{11} &= D_{11} + 2\pi\rho b U (c_0 - c_1 - c_3), \\
C_{12} &= D_{12} + \left[1 + (c_0 - c_1 - c_3)(1-2a)\right]\pi\rho b^2 U, \\
C_{13} &= D_{13} + \left[T_{11}(c_0 - c_1 - c_3) - T_4\right]\rho U b^2, \\
C_{14} &= 2\pi\rho U^2 b (c_1 c_2 + c_3 c_4), \\
C_{21} &= D_{21} - 2\pi\rho b U\left(a + \frac{1}{2}\right)(c_0 - c_1 - c_3), \\
C_{22} &= D_{22} + \left(\frac{1}{2} - a\right)\left[1 - (c_0 - c_1 - c_3)(1+2a)\right]\pi\rho b^3 U, \\
C_{23} &= D_{23} + \left(T_1 - T_8 - (c-a)T_4 + \frac{T_{11}}{2}\left(a + \frac{1}{2}\right)T_{11}(c_0 - c_1 - c_3)\right)\rho b^3 U, \\
C_{24} &= -2\pi\rho U^2 b^2\left(a + \frac{1}{2}\right)(c_1 c_2 + c_3 c_4), \\
C_{31} &= D_{31} + \rho U b^3 T_{12}(c_0 - c_1 - c_3), \\
C_{32} &= D_{32} + \left[T_4\left(a - \frac{1}{2}\right) - T_1 - 2T_9 + T_{12}\left(\frac{1}{2} - a\right)(c_0 - c_1 - c_3)\right]\rho U b^3, \\
C_{33} &= D_{33} + \frac{(T_1 T_{12}(c_0 - c_1 - c_3) - T_4 T_{11})\rho U b^3}{2\pi}, \\
C_{34} &= \rho U^2 b^2 T_{12}(c_1 c_2 + c_3 c_4), \\
C_{41} &= -\frac{1}{b}, \qquad C_{42} = a - \frac{1}{2}, \qquad C_{43} = -\frac{T_{11}}{2\pi}, \qquad C_{44} = \frac{(c_2 + c_4)U}{b}.
\end{aligned}$$

The stiffness matrix is

$$K = \begin{bmatrix} K_{11} & K_{12} & K_{13} & K_{14} \\ K_{21} & K_{22} & K_{23} & K_{24} \\ K_{31} & K_{32} & K_{33} & K_{34} \\ K_{41} & K_{42} & K_{43} & K_{44} \end{bmatrix},$$

with

$$\begin{aligned}
K_{11} &= m_w \omega_h^2 k_{h0}, \\
K_{12} &= 0, \\
K_{13} &= 2\rho U^2 b (c_0 - c_1 - c_3), \\
K_{14} &= 2\rho U^3 (c_2 c_4)(c_1 + c_3), \\
K_{21} &= 0, \\
K_{22} &= m_w \omega_\alpha^2 r_\alpha^2 b^2 k_{\alpha 0}^2 - 2\pi \left(\frac{1}{2} + a\right)(c_0 - c_1 - c_3)\rho b^2 U^2, \\
K_{23} &= \rho U^2 b^2 \left[ T_4 + T_{10} - T_{10}(2a+1)(c_0 - c_1 - c_3) \right], \\
K_{24} &= -2\pi \rho U^3 b^2 \left(a + \frac{1}{2}\right)(c_1 c_2 + c_3 c_4), \\
K_{31} &= 0, \\
K_{32} &= \rho U^2 b^2 T_{12} (c_0 - c_1 - c_3), \\
K_{33} &= m_w \omega_\beta^2 r_\beta^2 b^2 k_{\beta 0}^2 + \frac{(T_5 - T_4 T_{10} + T_{10} T_{12}(c_0 - c_1 - c_3))\,\rho U^2 b^2}{\pi}, \\
K_{34} &= \rho U^3 b T_{12} \left(c_2 c_4 (c_1 + c_3)\right), \\
K_{41} &= 0, \qquad K_{42} = \frac{U}{b}, \qquad K_{43} = \frac{T_{10} U}{b\pi}, \qquad K_{44} = \frac{c_2 c_4 U^2}{b^2}.
\end{aligned}$$

Here $D_{ij}$ denote the entries of the structural damping contribution. In the computations reported in the paper, these entries are generated from the chosen Rayleigh damping model[25].